\documentclass[twocolumn]{aastex631}

	\usepackage{url}
	\usepackage{amssymb}		
	\usepackage{graphicx}		
	\usepackage{hyperref}		

	\usepackage{xcolor} 		

	\usepackage{amsmath} 
	\usepackage{scalerel} 

	\usepackage{color} 
	\usepackage{soul} 
	\usepackage{comment} 

	\usepackage{soul} 

	\newcommand{\muram}{{MURaM}} 

	\newcommand\avg[1]{$\left\langle \text{#1} \right\rangle$} 

	\makeatletter
	\newcommand*\bigcdot{\mathpalette\bigcdot@{.5}} 
	\newcommand*\bigcdot@[2]{\mathbin{\vcenter{\hbox{\scalebox{#2}{$\m@th#1\bullet$}}}}} 
	\makeatother

	\newcommand\vectr[1]{$\mathbf{#1}$}

	\newcommand{\Fm}{$\mathbf{F}$} %

	\newcommand{\dPev}{$\Delta \mathbf{P_e}$} %
	\newcommand{\Pev}{$\mathbf{P_e}$} 
	\newcommand{\Pe}{$\text{P}_{\text{e}}$} 
	
	\newcommand{\fRFPem}{$\mathbf{R}_{\text{P}_{\text{e}}}$} %

	\newcommand{\AKPev}{$\mathbf{A}_{\text{P}_{\text{e}}}$} %

	\newcommand{\los}{line-of-sight} 
	\newcommand{\Vlosv}{$\mathbf{V_{los}}$} 
	\newcommand{\Vlos}{V$_{\text{los}}$} 
	\newcommand{\fRFVlosm}{$\mathbf{R}_{\text{V}_{\text{los}}}$} 
	\newcommand{\AKVlosv}{$\mathbf{A}_{\text{V}_{\text{los}}}$} 
	\newcommand\errv{$\boldsymbol{\varepsilon}$}

	\newcommand{\dSIv}{$\Delta \mathbf{I}$} 
	\newcommand{\SIv}{$\mathbf{I}$} 

	\newcommand{\Tv}{$\mathbf{T}$} 
	\newcommand\dTi[1]{$\text{$\Delta \text{T}$} (\tau_{#1})$} 
	\newcommand\Ti[1]{T$(\tau_{#1})$} 
	\newcommand{\dTv}{$\Delta \mathbf{T}$} 
	\newcommand\fdTi[1]{$\frac{\text{$\Delta \text{T}$}}{\text{T}} (\tau_{#1})$} 

	\newcommand{\fdTv}{${\frac{\Delta \mathbf{T}}{\mathbf{T}}}$} 
	\newcommand{\fdTvflat}{${\Delta\mathbf{T}}/{\mathbf{T}}$} 
	\newcommand{\fdTvflati}{${\Delta\text{T}}/{\text{T}}$}
	\newcommand{\fdVlosvflat}{${\Delta \mathbf{V_{los}}}/{\mathbf{c_s}}$}
	\newcommand{\fdPevflat}{${\Delta \mathbf{P_e}}/{\mathbf{P_e}}$}
	
	\newcommand{\RFTm}{$\mathbf{R}_{\text{T}}^\prime$} 
	\newcommand{\fRFTm}{$\mathbf{R}_{\text{T}}$} 
	\newcommand{\fRFTmt}{$\mathbf{R}_{\text{T}}^\top$} %

	\newcommand{\AKTv}{$\mathbf{A}_{\text{T}}$} 
	\newcommand\AKTvi[1]{$\mathbf{A}_{\text{T}} (\tau_{#1} )$} 

	\newcommand{\GTFv}{$\mathbf{G}$} 

	\newcommand{\coeffv}{$\mathbf{C}$} 
	\newcommand{\coeffvt}{$\mathbf{C}^\top$} 
	\newcommand\coeffi[1]{$\text{c}_{\lambda_{#1}}$} 
		
	\newcommand\logtau{log$\ \tau$}
	\newcommand\pinv{pseudo-inverse} 

	\newcommand{\ctAKPev}{$\mathbf{A}_{\text{P}_{\text{e}}}$} %
	\newcommand{\ctAKVlosv}{$\mathbf{A}_{\text{V}_{\text{los}}}$} %
	
\shorttitle{iterOLA-paper1}
\shortauthors{Agrawal et al.}

\graphicspath{{./}{figures/}}

\begin{document}

\title{An iterative OLA method for inversion of solar spectropolarimetric data:  I.  Single and multiple variable inversions of thermodynamic quantities}
\author[0000-0001-6514-8944]{Piyush Agrawal}
\affiliation{Department of Astrophysical and Planetary Sciences, University of Colorado, Boulder, CO 80309, USA} 
\affiliation{Laboratory for Atmospheric and Space Physics, University of Colorado, Boulder, CO 80303, USA}
\affiliation{Southwest Research Institute, Boulder, CO 80302, USA}

\author[0000-0002-9232-9078]{Mark P. Rast}
\affiliation{Department of Astrophysical and Planetary Sciences, University of Colorado, Boulder, CO 80309, USA} 
\affiliation{Laboratory for Atmospheric and Space Physics, University of Colorado, Boulder, CO 80303, USA}

\author[0000-0001-9550-6749]{Basilio Ruiz Cobo}
\affiliation{Instituto de Astrofisica de Canarias, La Laguna, Tenerife, E-38200, Spain}
\affiliation{Departamento de Astrofísica, Univ. de La Laguna, La Laguna, Tenerife, E-38205, Spain}

\begin{abstract}
		This paper describes an adaptation of the Optimal Localized Averaging (OLA) inversion technique, originally developed for geo- and helioseismological applications, to the interpretation of solar spectroscopic data. It focuses on inverting the thermodynamical properties of the solar atmosphere assuming that the atmosphere and radiation field are in Local Thermodynamic Equilibrium (LTE).  We leave inversions of magnetic field and non-LTE inversions for future work.  The advantage with the OLA method is that it computes solutions that are optimally resolved (in depth) with minimal cross-talk error between variables. Additionally, the method allows for direct assessment of the vertical resolution of the inverted solutions.  The primary challenges faced when adapting the method to spectroscopic inversions originate with the possible large amplitude differences between the atmospheric model used to initiate the inversion and the underlying atmosphere it aims to recover, necessitating the development of an iterative scheme.  Here we describe the iterative OLA method we have developed for both single and multivariable inversions and demonstrate its performance on simulated data and synthesized spectra. We note that when carrying out multivariable inversions, employing response function amplification factors can address the inherent spectral-sensitivity bias that makes it hard to invert for less spectrally-sensitive variables.  The OLA method can, in most cases, reliably invert as well as or better than the frequently employed Stokes Inversion based on Response functions (SIR) scheme, however some difficulties remain. In particular, the method struggles to recover large-scale offsets in the atmospheric stratification. We propose future strategies to improve this aspect. 
	
\end{abstract}

	
\section{Introduction}

	Solar physicists often rely on inversion methods to infer the physical properties of the solar atmosphere from spectroscopic (or spectropolarimetric) measurements.  Photon interactions with the plasma alter their intensity and polarization state, allowing recovery of the properties of the atmosphere through which they have traveled.  Depth dependent inference of the physical properties is possible because the opacity changes with wavelength, dramatically across spectral lines, providing varying sensitivity to different regions of the atmosphere. 

	Over the years, robust methods have been developed to model the spectral output of the solar plasma (the forward problem) or invert for the atmospheric properties from observed spectrum (the inverse problem).  These employ a range of approximations
		\citep{mihalas78, sir96, navarro01, josecarlos_book, luis06, ramos12, sirLR16, rodriguez17}: 
	Milne-Eddington (e.g.
		\cite{me_unno56, harvey72, auer77, landolfi84, skumanich85, me_sku87, suarez07a, borrero11c, me_centeno14, Yadav2017}), 
	Local Thermodynamic Equilibrium (LTE, e.g. 
		\cite{mihalas78, sir92, solanki92, frutiger98, furtiger00, navarro01, gray2005}), 
	Non-local Thermodynamic Equilibrium (e.g. 
		\cite{mihalas78, navarro00, ramos08, navarro15, rodriguez16, milicRF, milic18, stic19, desire22}). %
	In general, inversion problems are underdetermined and one must carefully consider uniqueness, but within the model assumptions, solar inversions have yielded valuable information about the morphology of sunspots (e.g.
		\cite{collados94, solanki03, suarez05, borrero11, borrero11b, borrero19, borrero21}, 
	properties of quiet-sun magnetism (e.g.
		\cite{stenflo82, bueno04, lites08, gonzalez08, suarez07, stenflo10, rubio19, arjona21}),
	and multi-thermal structure of the solar chromosphere (e.g.	
		\cite{ramos08, casini09, bueno10, priest18, yadav20, anan21, morosin22}), as examples.

	In this paper, we examine an inversion approach that has not previously been applied to the spectroscopic data. We invert synthetic spectra for the physical properties of a simulated solar atmosphere using the Optimal Localized Averaging (OLA) technique.  OLA was originally developed in the geoseismological context  to invert for the Earth's interior structure~\citep{backus67, backus68, backus70} and has been used with great success in helioseismology to infer the interior structure and dynamics of the Sun based on its acoustic oscillations~\citep{gough1982, gough1985, jeffrey1988, dalsgaard1990, pijper92, olarls1993, thompson94, basu2016}.  
	The primary advantage of the OLA method lies in its ability to carefully constrain, given an observed spectrum, the depth dependance of the inverted solution and selectively invert where optimally localized depth averaged solutions can be obtained. Along with the inverted solution, the method provides a clear measure of the resolution achievable using a given spectrum and an assessment of the minimal smoothing required for a reliable solution at that depth.	

	As a first step, this paper focuses on the use of OLA for the inversion of spectroscopic data (Stokes \SIv) to determine the thermodynamic properties of the solar photosphere, temperature \Tv, electronic pressure \Pev, and the \los\ velocity \Vlosv\ as functions of depth, assuming that the spectrum is formed under conditions for which the LTE approximation is valid.  Note that throughout this paper bold letters correspond to vectors (or matrices), so that \SIv\ indicates a vector of the intensity at wavelengths $\lambda$, while \Tv, \Pev, and \Vlosv\ represent vectors of the atmospheric properties as a function of optical depth log $\tau$. 
		Note that throughout this article, optical depth is evaluated in the continuum at 500nm. 
	For spectroscopic inversions, adaptations of the OLA method are required to allow inversion for non-linear perturbations (the initial guess atmospheric model on which the inversion is based, may not be close to the underlying atmosphere it aims to recover).  We discuss those adaptations and compare the inversion results we achieve to those obtained using the Stokes Inversion based on Response functions (SIR) inversion method~\citep{sir92, sir96, josecarlos_book, sirLR16}.  
	
	In Section~\ref{intro_general}, we provide a brief introduction to spectroscopic inversions, discuss the problems faced and some limitations of current inversion approaches. In Sections~\ref{sec2_olaintro} and~\ref{sec_multivar}, we describe the OLA method we have developed and its application to single and multivariable inversions.  We compare the OLA inversion output directly to the simulated atmospheres used to construct the synthetic spectrum and assess the method's reliability.  Finally, we summarize and discuss future prospects in Section~\ref{sec_conclusion}.

\section{Brief introduction to spectroscopic inversions}
\label{intro_general}
		\begin{figure*}[!t] 		
		    \begin{center}
			\includegraphics[width=.98\textwidth]{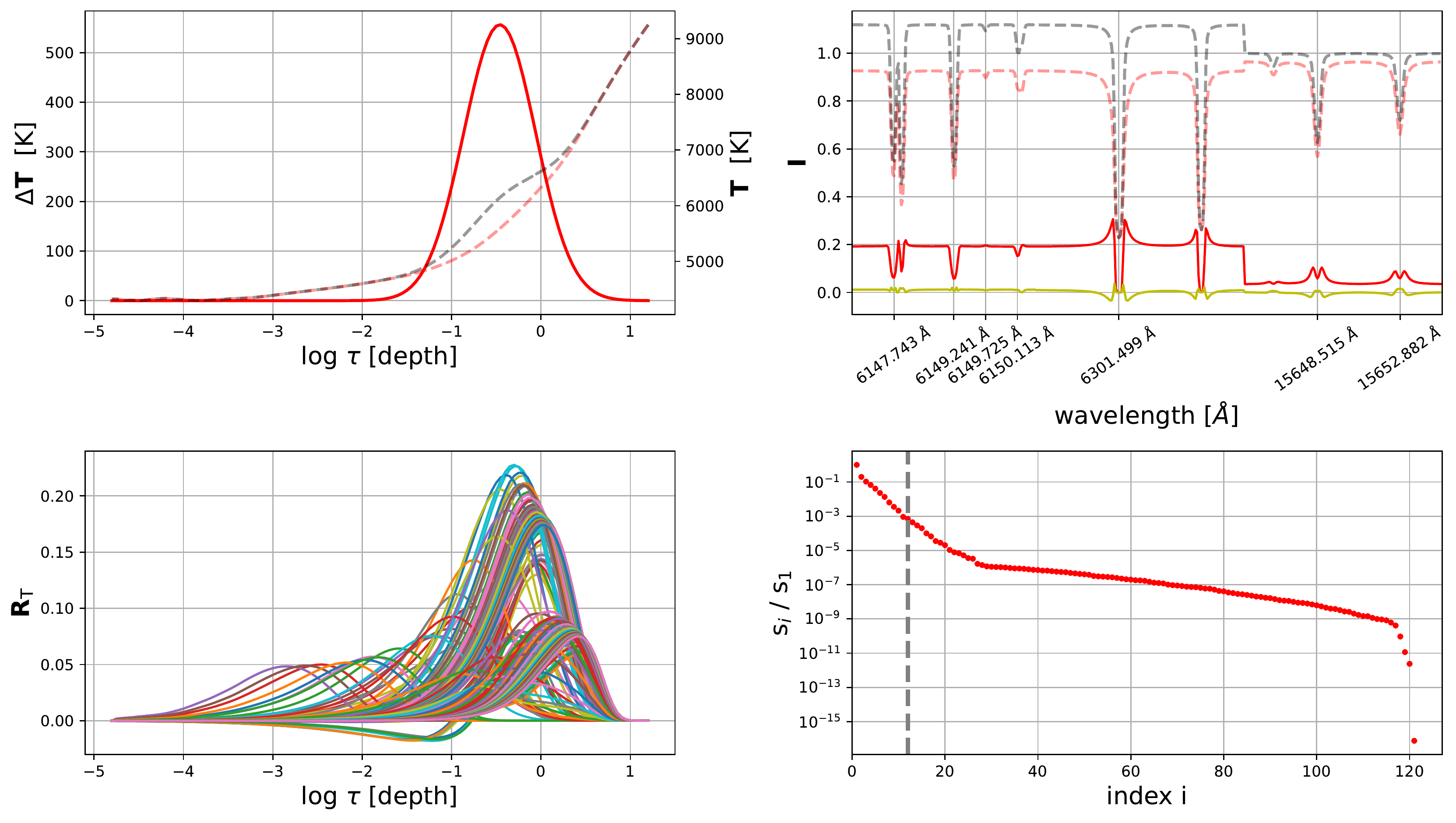}
		    \end{center} 
		    \caption{({Top-left}) An example test case where underlying \Tv\ (gray-dashed) is constructed by adding a Gaussian-shaped perturbation \dTv\ (red-solid, plotted with a different scaling) peaked at \logtau\ $= -0.50$ (width = $0.40$ $\Delta$\logtau, perturbation amplitude $=10\%$) to the mean \muram\ guess \Tv\ (red-dashed). ({Top-right}) Corresponding spectrum \SIv\ (dashed curves) and spectral-differences \dSIv\ (red-solid curve) for spectral lines mentioned in Table~\ref{table_linelist}, synthesized using SIR forward solver.  Intensities are normalized using the HSRA model \citep{hsra}.  Overplotted in yellow are the higher order errors, given by $=$ $\Delta \mathbf{I}\ -$ \fRFTm$^\top \bigcdot$ \fdTv.		    
		    ({Bottom-left}) Fractional temperature response functions as a function of optical depth (log $\tau$ at 500nm) for individual wavelengths that make up the observed spectrum, computed for mean \muram\ guess model using SIR. 
		    ({Bottom-right}) Singular values of the \fRFTm\ matrix, normalized using the largest singular value $s_1$. Vertical grey-dashed line corresponds to index $i=12$ for which dominant singular values cumulatively add to $99.9\%$ of the total sum. \\}    
		    \label{fig4_1_testcase1_model_and_spectra}				
		\end{figure*}

	To begin, we briefly review spectroscopic inversion strategy, defining the notational scheme we employ and focusing on those aspects of inversion which are critical to the development of both the single and multivariable OLA inversions we develop in Sections~\ref{sec2_olaintro} and~\ref{sec_multivar} respectively.
	
	As an illustrative example, consider the single variable inversion of spectroscopic data (Stokes \SIv) for the plasma temperature \Tv.  In general, spectroscopic inversions are carried out by starting with an initial guess for \Tv\ and then solving for the difference \dTv\ between it and the actual (or underlying, observed) \Tv\ that accounts for the spectral difference \dSIv\ between the observed spectra and that of the guess model. When the inversion is complete, if the difference between the spectra derived from the inverted model and that observed falls within some specified tolerance, the inverted model is taken to approximate the underlying atmosphere. However, due to the underdetermined nature of the problem, a good spectral fit doesn't guarantee recovery of the underlying atmosphere.  Most fundamentally this is because the radiative transfer equation is an integral equation, the observed intensity is a weighted integral of the physical properties of the atmosphere over a range of depths, and thus different atmospheric stratifications are consistent with a given observed spectra (and spectral differences \dSIv).  

	A schematic of the canonical spectral inversion problem is shown in Figure~\ref{fig4_1_testcase1_model_and_spectra}.  In the top-left panel, the underlying \Tv\ profile is shown with the gray-dashed curve and the guess \Tv\ with the red-dashed curve.  The difference \dTv\ is overplotted (with a different scaling) as red-solid curve.  Neither the underlying \Tv\ (gray-dashed curve) or the difference \dTv\ (red-solid curve) are known ahead of the inversion, and it is the \dTv\ profile that the inversion aims to capture. The corresponding spectra and spectral differences (for the spectral lines used in this work, see Table~\ref{table_linelist}), as synthesized using the SIR forward solver, are shown in the top-right plot of the figure.  The gray-dashed spectrum is what would be observed if the true underlying atmosphere were to be observed and the red-dashed spectrum is the spectrum of the guess model atmosphere. The inversion itself aims to account for their difference (red-solid curve). 
		
	One way to proceed is to linearize the relationship between \dSIv\ and \dTv\ ~\citep[e.g.,][]{josecarlos_book}).   This relationship can be written as a 1st order linear system of equations,
			\begin{eqnarray} \label{eqn2_1storder1var1}
				\Delta \mathbf{I} \  &=& \  \mathbf{R}_\text{T}^{\prime\top} \bigcdot \mathbf{\Delta{T}}  \  + \  \boldsymbol{\varepsilon}. 
			\end{eqnarray}

	\noindent
	When discretized in wavelength and optical depth, \RFTm\ is a matrix containing the linear temperature response function at all optical depths and wavelengths under consideration 
		~\citep{landi77, sir92}.  
		This is computed for the guess model, and $\mathbf{R}_\text{T}^{\prime\top} $ in Equation~\ref{eqn2_1storder1var1} denotes its transpose. The temperature response function captures the spectral sensitivity $\partial \text{I}_\lambda$, at a given wavelength, to an infinitesimal change in temperature $\partial \text{T}_\tau$ at each optical depth, $\text{R}_\text{T}^{\prime} (\tau, \lambda) = \partial \text{I}_\lambda/\partial \text{T}_\tau$. 
	In Equation~\ref{eqn2_1storder1var1}, \errv\ captures both observational and instrumental errors and contributions from higher order terms in the relationship between \dSIv\ and \dTv\ (those not included in the linear response function $\text{R}_\text{T}^{\prime} (\tau, \lambda)$).  When the guess model is far from the observed atmosphere, \dTv\ is not small, and the relationship between \dSIv\ and \dTv\ is no longer linear.  Under these conditions, the missing higher order terms (order $\left({\mathbf{{\Delta T}}}\right)^2$ and higher) contribute significantly to \errv.  The contribution of higher order terms is indicated, for our example problem, by the yellow curve in top-right plot in Figure~\ref{fig4_1_testcase1_model_and_spectra}. In this work, we do not include any observational or instrumental noise, but their effect on the inversion solution is discussed in Section~2.2.7 of \cite{ourthesis}.  We also ignore error due to discretization or numerical round-off, as these are typically smaller than observational/instrumental noise. 

	We note that, in this paper, unlike in most previous applications, we employ fractional response functions, $\mathbf{R}_\text{T}$, equal to the ratio of change in intensity at a given wavelength to the fractional change in temperature at a given depth.  Thus Equation~\ref{eqn2_1storder1var1} becomes,
		\begin{eqnarray} \label{eqn2_1storder1var}
			\Delta \mathbf{I}\ &=& \  \mathbf{R}_\text{T}^\top \bigcdot \frac{\Delta{\mathbf{T}}}{\mathbf{T}} \  + \  \boldsymbol{\varepsilon} .
		\end{eqnarray}

	\noindent	
	This removes the dimensional dependency of the spectral sensitivity to change in a given variable, facilitating comparison between the magnitudes of the response to different variables (see multivariable inversions in Section~\ref{sec_multivar}).  For the test case shown, the fractional temperature response functions are plotted in the bottom-left panel of Figure~\ref{fig4_1_testcase1_model_and_spectra}.  Each curve plots temperature response function for a different $\lambda$ as a function of log $\tau$, as computed using the SIR solver.

	Solving Equation~\ref{eqn2_1storder1var} for \fdTvflat\ requires computing the inverse of the \fRFTm\ matrix. This is challenging given that the system is ill-posed, as evident from the sharp decay of the singular values of the \fRFTm\ matrix (bottom-right plot in Figure~\ref{fig4_1_testcase1_model_and_spectra}) which reflects the limited independence of the response functions and thus the limited number of orthogonal modes available to represent \fRFTm. A full-rank matrix inverse (or \pinv) for an ill-posed system has large amplitude components which can amplify non-zero error \errv\ and result in an error dominated solution (\cite{hansen_dpc2, hansen94, hansen_rls2, numrecipe}).  It is critical to minimize \errv\ contributions so that the inverse solution largely reflects the underlying \fdTvflat\ difference rather than \errv\ in Equation~\ref{eqn2_1storder1var}. To prevent \errv\ dominated solutions, we rely on a lower rank \pinv\ (\fRFTm$^*)^{-1}$, for which $[$\fRFTm\ $\bigcdot$ (\fRFTm$^*)^{-1}]^\top$ is not strictly an identity matrix. This yields an inversion solution that at each depth has contributions from other depths (cross-talk error) and from other variables in multivariable inversions (see Section~\ref{sec_multivar}).  		

	We note that this is true of all inversions, not particular to the OLA method we discuss in this paper. 
	It is in general not possible to minimize cross-talk error without amplifying the error due to non-linearity or the observations, as eliminating cross-talk requires employing a larger rank \pinv\ matrix while minimizating \errv\ requires a smaller rank pseudo-inverse. The goal is to regularize the solution so that it 'optimally balances' these error contributions. Regularization, irrespective of the inversion scheme employed, effectively removes the smaller magnitude singular-value contributions to the \pinv\ matrix
		\citep{pinv2, hansen_dpc2, hansen94, hansen_rls2, numrecipe}.  
	A regularized solution is typically one that is a smoother version of the underlying \fdTvflat, one that is cross-talk dominated and can thus fail to recover sharp gradients that may be present.   On the other hand, a more weakly regularized solution, obtained by keeping smaller singular values, is able to recover sharper gradients present, but is also more likely to be \errv\ dominated and highly oscillatory with depth. 
	Determining the appropriate degree of regularization (what rank inverse to employ) to optimally balance resolution (minimize cross-talk) and \errv\ amplification is one of the most challenging aspects of inversions. This is because the spectral difference metric by which we assess how close the inverted solution is to the actual perturbation is degenerate. Different degrees of regularization can lead to significantly different solutions all with similarly good fits to the observed spectrum.

	Current state of the art spectropolarimetric inversion codes, such as SIR, obtain a smooth solution by solving for the underlying perturbations at a limited number of user-defined depth locations called 'nodes.' The solution for other depth points are interpolated using the inverted nodal values. Inverting at limited depths indirectly reduces the \pinv\ matrix rank when solving Equation~\ref{eqn2_1storder1var}, and thus minimizes \errv\ contributions.  The resulting solutions are therefore typically cross-talk error dominated, especially in the shallower and deeper regions where response function sensitivities are less orthogonal. Importantly, the solutions do not typically correspond to the 'most vertically resolved solution possible' given the set of spectral response functions available.  It is very hard, using this approach, to know a priori what gradients are recoverable and how many nodes to employ, given the observed spectrum.  

	The OLA method was developed in the geo- and helioseismic communities to address these difficulties. Over the remainder of this paper, we adapt and apply the method to spectropolarimetric inversions, 'optimally' minimizing cross-talk and \errv\ error contributions to the inverted solution at each depth. This allows solutions at the vertical-resolution limit, given observations of specific spectral content.  		

\section{OLA single variable inversion} \label{sec2_olaintro}
	In this section we describe the OLA methology for single variable, \fdTvflat, inversions using an artificially constructed test case.  The  test case, adds a depth dependent temperature perturbation $\Delta\bf{T}$ to the horizontal mean state of a \muram\ simulation (\cite{voegler05, rempel14}) to construct the underlying \Tv\ model.  Starting with the mean \muram\ model as the initial guess, we conduct a OLA inversion for \fdTvflat\ and compare the deduced $\Delta\bf{T}$ with that imposed. This allows an assessment of the capabilities and limitations of the OLA method and illustrates the implementation changes necessary in the spectropolarimetric context. Since we focus on inverting the spectral data for temperature perturbations in the absence of other variables, no perturbations are added to \vectr{P_e} of the underlying mean \muram\ model and \vectr{V_{los}} is taken to be zero. 

	Development of the OLA spectral inversion technique proceeds as follows: using the forward spectral synthesis capabilities of SIR, we compute the synthetic spectrum of the observed and the guess models (Figure~\ref{fig4_1_testcase1_model_and_spectra}, top-right plot) and  the fractional spectral response function matrix \fRFTm\ using the initial guess model (Figure~\ref{fig4_1_testcase1_model_and_spectra}, bottom-left plot). The computed response functions, along with the spectral difference \dSIv\ between observed and guess spectra, define the 1st order linear system of Equations~\ref{eqn2_1storder1var} which must be solved for \fdTvflat. Once obtained, this inverted \fdTvflat\ profile can be compared to the underlying \fdTvflat\ for accuracy and reliability.

	\subsection{Averaging kernels} \label{chp4_2_akconst}
			\begin{figure}[!t]		
			    \begin{center}
				\includegraphics[width=0.45\textwidth]{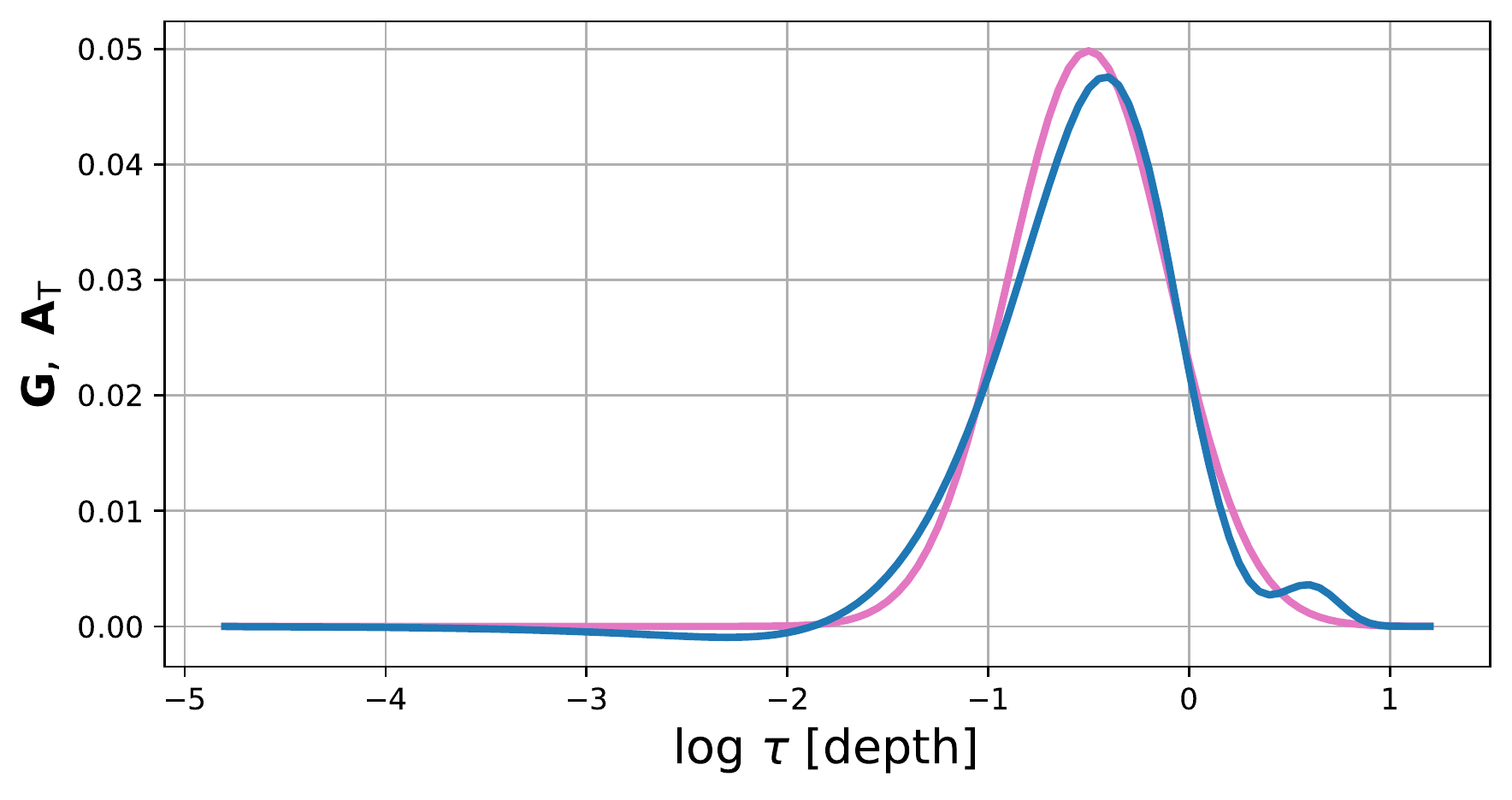}
			    \end{center} 
			    \caption{Gaussian target function $\bf G$ with peak at \logtau\ $= -0.5$, and width $=0.4$ $\Delta$ log $\tau$ plotted as pink curve. Blue curve plots the averaging kernel  \AKTv\ obtained by solving Equation~\ref{eqn_akconstruct} using a rank $k=5$ \pinv\ of \fRFTm\ (comprised of the response functions plotted in the lower left panel of Figure~\ref{fig4_1_testcase1_model_and_spectra}). \\ }
			    \label{fig4_2a:akpictorial}	
			\end{figure}

		The OLA method is fundamentally different from SIR which aims to find an overall smooth solution that best fits the observed spectra by simultaneously solving for the underlying perturbation at all nodal depths.  With OLA, inversion for the underlying perturbation proceeds one depth at a time, with the goal of finding the 'best possible' solution at that depth, the solution that optimally minimizes contributions from both \errv\ and from other depths (cross-talk). This is the optimally localized solution. 
		
		To obtain an OLA inversion at a target depth $\tau_i$, we first determine the coefficients \coeffv\ = [\coeffi{1}, \coeffi{2}, \coeffi{3}, ...]$^\top$, whose inner product with the response function matrix results in an averaging kernel that best mimics a user defined target function. This coefficient vector, \coeffv\ is then used with \dSIv\ to obtain the inverted solution at depth $\tau_i$. In other words, the averaging kernel \AKTvi{i}$\ =\mathbf{R}_\text{T} \bigcdot \mathbf{C}$, is obtained by solving the linear system of equations:
			\begin{eqnarray} \label{eqn_akconstruct} 
				\mathbf{R}_\text{T} \ \bigcdot \ \mathbf{C} \ &=& \ \mathbf{G} (\tau_i, \sigma) ,
			\end{eqnarray}

		\noindent
		for \coeffv, where \GTFv\ is a user-defined target function, localized at the target depth $\tau_i$.  The target function is usually taken to be a normalized Gaussian centered at $\tau_i$.  An example is shown in Figure \ref{fig4_2a:akpictorial}. Plotted in pink is a Gaussian target function with peak at \logtau\ $= -0.5$ and width $=0.4$ (in $\Delta$ \logtau\ units), and in blue the recovered \AKTv\ when solving Equation~\ref{eqn_akconstruct} with \pinv\ matrix rank $k=5$. The rank was chosen based on dominant singular values of \fRFTm, those that add up to $95\%$ of the total sum. 

		Once computed, the coefficients \coeffv\ can be used to obtain the inverted solution at depth $\tau_i$.  Taking the dot product of Equation~\ref{eqn2_1storder1var} with \coeffv$^\top$ and recognizing $\mathbf{R_\text{T}} \bigcdot \mathbf{C}$ as the averaging kernel \AKTvi{i} yields 
			\begin{eqnarray} \label{eqn4_2b:olainv} 
				\mathbf{C^\top} \ \bigcdot \  \mathbf{\Delta I} \  &=& \  \mathbf{A}_\text{T}^\top (\tau_i) \bigcdot \ \mathbf{\frac{\Delta T}{T}} \ + \ \mathbf{C^\top} \bigcdot \boldsymbol{\varepsilon}  .
			\end{eqnarray} 

		\noindent
		Here, \coeffv$^\top \ \bigcdot$ \dSIv\ is the {\it inversion solution at} $\tau_i$, which, if the error contribution \coeffv $^\top \bigcdot$ \errv\ is small, corresponds to an optical depth averaged value of the underlying \fdTvflat, defined as \avg{\fdTi{i}}  $\equiv$ $\mathbf{A}_\text{T}^\top (\tau_i)\  \bigcdot$ \fdTv.  The inverse solution for \dTi{i} is the product, guess \Ti{i} $\times$ (\coeffv$^\top \ \bigcdot$ \dSIv) $\approx$ guess \Ti{i} $\times$  \avg{\fdTi{i}}.
		The narrower the averaging kernel, the closer \avg{\fdTi{i}} lies to the local \fdTi{i}; if a $\delta$-function averaging kernel could be constructed \avg{\fdTi{i}} $\equiv$  \fdTi{i}. 

		The minimum kernel width is, however, determined by the response function set available and the \pinv\ matrix rank employed.   Equations~\ref{eqn_akconstruct}, like Equations~\ref{eqn2_1storder1var}, represents an ill-posed system. Constructing a narrower \AKTv\ requires employing larger rank $k$, which typically results in larger amplitude terms in \coeffv.  These larger amplitude terms can then amplify \errv\ through $\mathbf{C^\top} \bigcdot \boldsymbol{\varepsilon}$ in Equation~\ref{eqn4_2b:olainv} and lead to an inverted solution \coeffv$^\top \ \bigcdot$ \dSIv\ that is \errv\ dominated. While there is an upper-limit to the magnitude of \AKTvi{i}$^\top \bigcdot$ \fdTv, which depends on the magnitude of underlying \fdTvflat, no such limit exists for $\mathbf{C^\top} \bigcdot \boldsymbol{\varepsilon}$.  
		
		In summary, when inverting at depth $\tau_i$, the goal is to construct as narrow an averaging kernel as possible while preventing the error due to non-linearity and observational error (\errv\ in Equation~\ref{eqn4_2b:olainv})  from dominating the solution.
		The kernel width is limited by orthogonality of the response functions included in \fRFTm\ and the \pinv\ matrix rank $k$ employed. While the orthogonal sensitivities depend on both the guess model and the spectrum observed, rank is user-defined and determines the amount of regularization. As previously mentioned, a larger $k$ allows for narrower \AKTv\ construction but typically results in \errv\ dominated solutions. A smaller rank only allows wider \AKTv\ construction, which then produces a cross-talk error dominated solution, and the challenge is to optimally balances these error regimes. 	
		
		In Section~\ref{subsec_iter_OLA}, we discuss an iterative scheme for OLA recovery of non-linear perturbations which includes a method for determination of rank for each iteration.  In the next section, we discuss how the OLA inversions are carried out over multiple optical depths once the rank $k$ has been chosen.
		
	\subsection{OLA inversion at multiple depth locations} \label{sec_olainv_ata_depth}
			\begin{figure}[!tp] 
				\begin{center}
				\includegraphics[width=0.48\textwidth]{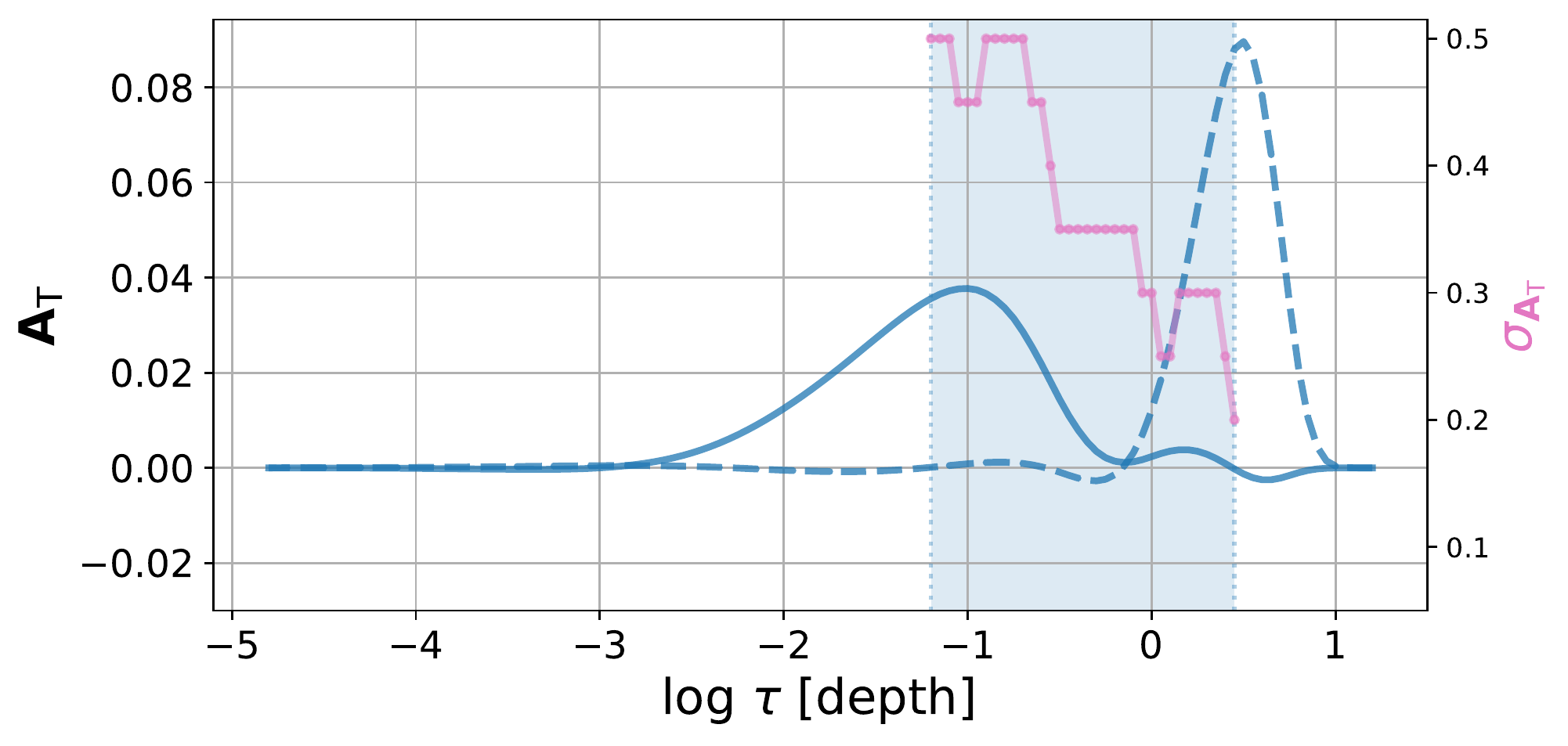}
			    \end{center} 
			    \caption{Blue-shaded region corresponds to the depth range over which OLA temperature inversion can be achieved (OLA inversion window), when using rank $k=5$ \pinv\ of \fRFTm\ (computed for mean \muram\ guess model, bottom-left plot in Figure~\ref{fig4_1_testcase1_model_and_spectra}). The corresponding minimum kernel widths $\sigma_{\mathbf{A}_\text{T}}$ that satisfy the target function fitting constraints of Section~\ref{sec_olainv_ata_depth} are shown in pink. Example averaging kernels at the edges of the inversion window are plotted in blue and blue-dashed. \\ }
			    \label{fig_sir_AKs}
			\end{figure}

		To obtain the optimally vertically-resolved inversion at $\tau_i$, we need to determine the narrowest averaging kernel \AKTv\ that can be constructed at that depth. For this, we define a range of target function widths spanning from 0.05 to 0.5, in steps of 0.05 (model grid spacing in $\text{log}\ \tau$ units). Starting with the minimum-width target function \GTFv, normalized to have area one and centered on the target depth $\tau_i$, we solve for coefficients \coeffv\ given the \pinv\ rank. We then compare the averaging kernel (\fRFTm\ $\bigcdot$\ \coeffv) achieved to the target function for goodness-of-fit.  
		We chose a fit measure based on the L1 norm of the difference between \GTFv\ and \AKTv, the sum of absolute-difference between the two, as it approximately corresponds to a $\%$ difference.  A goodness-of-fit criterion that is stringent (smaller tolerance to the difference) makes it harder to construct averaging kernels (for a given rank $k$), while a relaxed criterion (larger tolerance) can result in averaging kernels that only very poorly resemble the target function. 	
		Empirically, we find that a L1 norm goodness-of-fit value of $0.2$ (mean $20\%$ difference) works well. 
		If the difference lies within this specified upper-limit, we conclude that the averaging kernel approximately represents the target function, and use the coefficients to compute the inverted solution \avg{\fdTi{i}} $= \mathbf{C}^\top \bigcdot \ \Delta \mathbf{I}$. Taking the inverted \dTi{i} to be guess T$(\tau_i) \times$\avg{\fdTi{i}}, the inversion is complete at location $\tau_i$, and the width of the target function approximates the vertical resolution of the inverted solution. The process is then repeated to invert at 'all possible' depths 
		using the same rank \pinv\ matrix. 
		When iterating (Section~\ref{subsec_iter_OLA}), we call this a single 'inversion cycle.' We note that, it is the averaging kernel width achieved at each depth, not the target function goal, that determines the depth sensitivity of the solution.

		For some depths, it is possible that an averaging kernel can not be constructed to fit even the widest target function with the specified fit tolerance. In such cases we set the inverted \fdTvflati\ to zero and conclude that an OLA inversion cannot be achieved at that location, given the rank $k$ employed and response function set available. This failure often occurs above and below a limit range of depths, as the orthogonal sensitivities of the response functions are non-uniformly distributed with depth. The depth range over which kernels can be constructed, which we termed the 'OLA inversion window,' is depicted by the blue-shaded region in Figure~\ref{fig_sir_AKs} for $k=5$ over the first inversion-cycle.  The narrowest kernels that can be constructed at the either end of the window are shown with blue-solid and blue-dashed curves, and the averaging kernel widths achieved at all depths, which approximates the resolution achieved, are shown in pink. 
			\begin{figure*}[!tp] 
				\begin{center}
				\includegraphics[width=\textwidth]{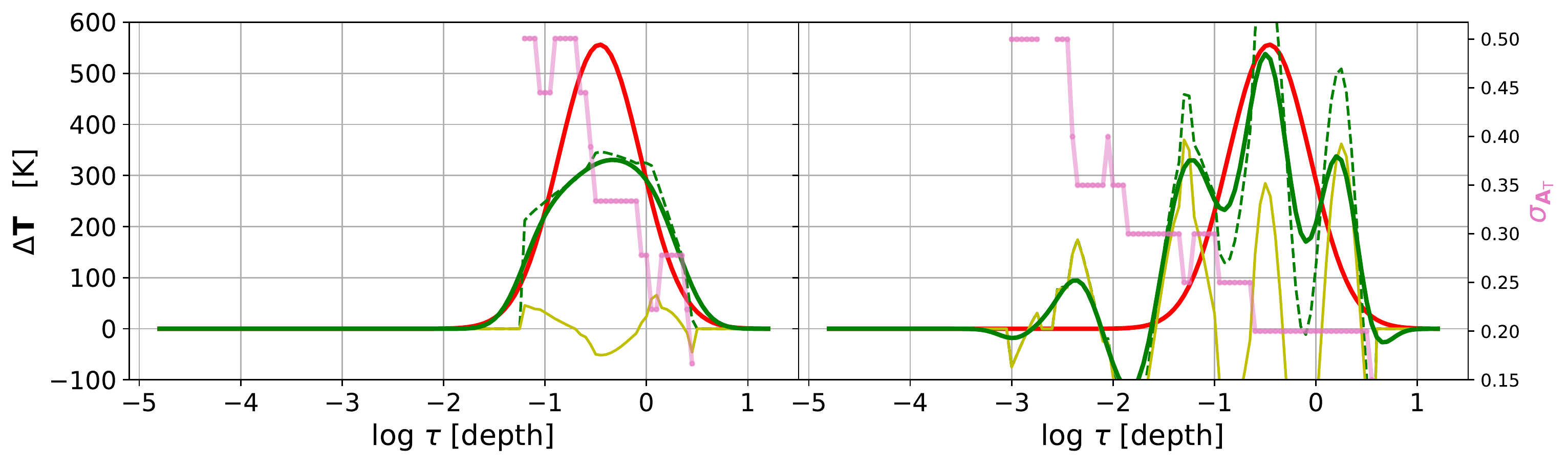}				
			    \end{center} 
			    \caption{OLA inversion for a temperature perturbation \dTv\ (red, replotted from the top-left panel in Figure~\ref{fig4_1_testcase1_model_and_spectra}) after the 1st inversion cycle when employing rank $k=5$ ({left plot}), and $k=10$ ({right plot}).  Green-dashed curves correspond to the un-smoothed results while green-solid curves correspond to a smoothed version of the inverted \dTv. The yellow curves plot the \errv\ error contributions (from \coeffv$^\top \bigcdot$\ \errv\ $\times$\ guess \Tv$(\tau_i)$) to inversions, and pink curve (plotted with a different scaling) corresponds to minimum kernel width achievable at each depth. \\ }	
			    \label{fig4_5_test1_lowhigh_rank_soln}
			\end{figure*}

		As with all inversions based on a limited spectrum, the inverted values at each depth are not independent (the averaging kernels are not $\delta-$functions). With OLA that dependance is made explicit via the depth-dependent averaging kernel profiles. While it is possible to compute the effective averaging kernels of a SIR inversion post hoc (see Appendix~\ref{Appendix_sirkernel}), the key difference between OLA and SIR is its ability to determine the minimum averaging kernel width possible at each depth and thus to invert only at depths where a localized kernel can be constructed. The latter is both a strength and a weakness. While providing a clear indication of where a solution is possible, the non-invertible regions pose a particular challenge when unknown perturbations exist outside the inversion window and iteration is required for the final solution (Section~\ref{subsec_iter_OLA}).  We discuss this 'edge-effect issue' in detail in Section~\ref{subsec4_4:largescale-dT}.  
		
		None-the-less, we note that the advantage of the OLA averaging kernel approach extends beyond the inversion itself. Even before performing an inversion, kernel construction allows direct quantitative assessment of spectral line combinations and their potential utility in inversions for a given variable at a given depth ({reference in preparation}). While that assessment also depends in detail on the starting guess model (via the response functions) and the \pinv\ rank $k$ ultimately employed, it can serve as a useful starting point for spectral line-combination determination best suited for a particular scientific inquiry. 
		
	\subsection{Iterative scheme} \label{subsec_iter_OLA} 

		When inverting spectropolarimetric data, the atmospheric profiles deduced after a single inversion cycle likely do not match the underlying solar profiles, and some spectral differences between the observed and model spectra persist. This is because the underlying perturbation magnitudes are usually large; the actual solar atmosphere is often quite far from the initial guess. The neglected higher order terms captured by \errv\ in Equation~\ref{eqn2_1storder1var} are not insignificant as the relationship between \dSIv\ and \fdTvflat\ is not linear. Using linear system of equations to recover non-linear perturbations in a single inversion cycle is typically not possible. 
		Employing a larger rank in the inversion results in \errv\ dominated solution, while a smaller rank is unable to recover steep gradients (if present).
		This is illustrated by Figure~\ref{fig4_5_test1_lowhigh_rank_soln} in which an OLA inversion results is shown for the test case (Figure~\ref{fig4_1_testcase1_model_and_spectra}, top-left plot) after one inversion cycle and for two different rank choices: $k=5$ and $k=10$. 

		Multiple inversion cycles are needed to arrive at an optimal solution. After each inversion cycle, the inverted model from the previous cycle is used as the starting guess for the next. Since the guess model changes with each cycle, the response functions must be recalculated at the start of each. If the iteration is convergent (the underlying \fdTvflat\ gets smaller with successive cycles), the non-linear contributions to \errv\ also decrease, and a higher rank \pinv\ matrix can be employed to refine the solution.  This is similar to what is done with SIR inversions when the number of nodes is increased as the inversion proceeds.  Here, we employ a smaller rank \pinv\ matrix during early cycles, to obtain a lower resolution solution which is less likely to be \errv\ dominated, and increase the rank as the inversion proceeds.  The resolution of the iterative inversion increases as the system approaches the linear approximation, with consequent lower total error and thus increased tolerance for larger-magnitude coefficients in \coeffv $^\top \bigcdot\ $\errv.  
		While using a lower rank pseudo-inverse matrix during early cycles likely results in cross-talk error dominated solutions, when these errors are made at depths where more localized kernels can be constructed, finer resolution solutions can be achieved with subsequent iterations.  The OLA scheme preserves, to some degree, the underlying \fdTvflat\ during early iterations and recovers it at higher resolution, where that is attainable in subsequent iterations, rather than losing it in a global best spectral-fit solution. 

		The goal is thus to determine the maximum rank $k$ to be used at the end of the full inversion-cycle series, which ultimately sets the best resolution achieved, and to formulate an algorithm that facilitates scaling up to it from the lower rank solutions earlier in the iteration cycle. 
		Note that the relative error after successive inversion cycles, both from cross-talk and \coeffv $^\top \bigcdot$ \errv, compared to the underlying perturbation magnitude remaining at each depth, determines both whether subsequent updates are convergent (underlying \fdTvflat\ gets smaller with cycle) and the maximum rank beyond which inversions are likely to be error dominated.  While there is an upper limit to the cross-talk error in \AKTv $^\top \bigcdot$ \fdTv, which depends on the magnitude of underlying \fdTvflat, error from $\mathbf{C^\top} \bigcdot \boldsymbol{\varepsilon}$\ is unbounded as it depends on the magnitude of the coefficients \coeffv, which gets larger with increasing rank. 
		
		The aim is to develop an iterative algorithm which itself determines the maximum rank that can be tolerated without knowing the actual solution to be achieved. The hope is that, as the system gets closer to the largest tolerable rank and the inversions become error dominated, this will be reflected in the spectral difference measure (how close the observed \SIv\ is to that obtained from the inverted model).  If that is the case,  one can conclude that the best solution has been achieved when the fit to the spectrum begins to worsen and that any further inversion cycles would lead to poorer result.  
		The error from \coeffv $^\top \bigcdot$ \errv\ is typically unbounded.  The magnitude of terms in \coeffv\ increase with the increasing \pinv\ rank over the iterative scheme we describe below.  If the non-linear contribution to \errv, which depend on the magnitude of \fdTvflat, decrease slower than the rate at which coefficient magnitudes increase, the solution will become error dominated.  
		Ideally, if solutions in a given iteration cycle become error dominated, that should be promptly reflected by a worsening of the inverted model spectral fits.  The iterative OLA inversion scheme would then have converged on the optimal solution, and iteration could be terminated without difficulty.  But, due to the underdetermined nature of the inverse problems, it is possible that there is a lag between when the inverted model fit gets worse and when the error amplification is reflected in the spectral difference.  
			
		In practice, we have developed the following scheme to minimize this possibility. We initiate the inversion by employing mean \muram\ model atmosphere as the starting guess model; we compute \SIv, \dSIv\ = observed $-$ guess \SIv, and the temperature response function matrix \fRFTm\ using it.  A smaller rank helps prevent \errv\ amplification, and we have empirically determined that a safe starting value can be based on dominant singular values that add to $95\%$ of the total singular-value sum. For our set of initial response functions, this yields a starting rank $k=5$. Using this rank, and the method described in Section~\ref{sec_olainv_ata_depth} above, we obtain an OLA inversion estimate of \fdTvflat.  This first inversion cycle solution, for the test case perturbation we are examining (red curves in Figures~\ref{fig4_1_testcase1_model_and_spectra} and \ref{fig4_5_test1_lowhigh_rank_soln}), is shown with a green-dashed curve in the left panel of Figure~\ref{fig4_5_test1_lowhigh_rank_soln}. 
		
		\begin{figure*}[!t] 
				\begin{center}
				\includegraphics[width=.98\textwidth]{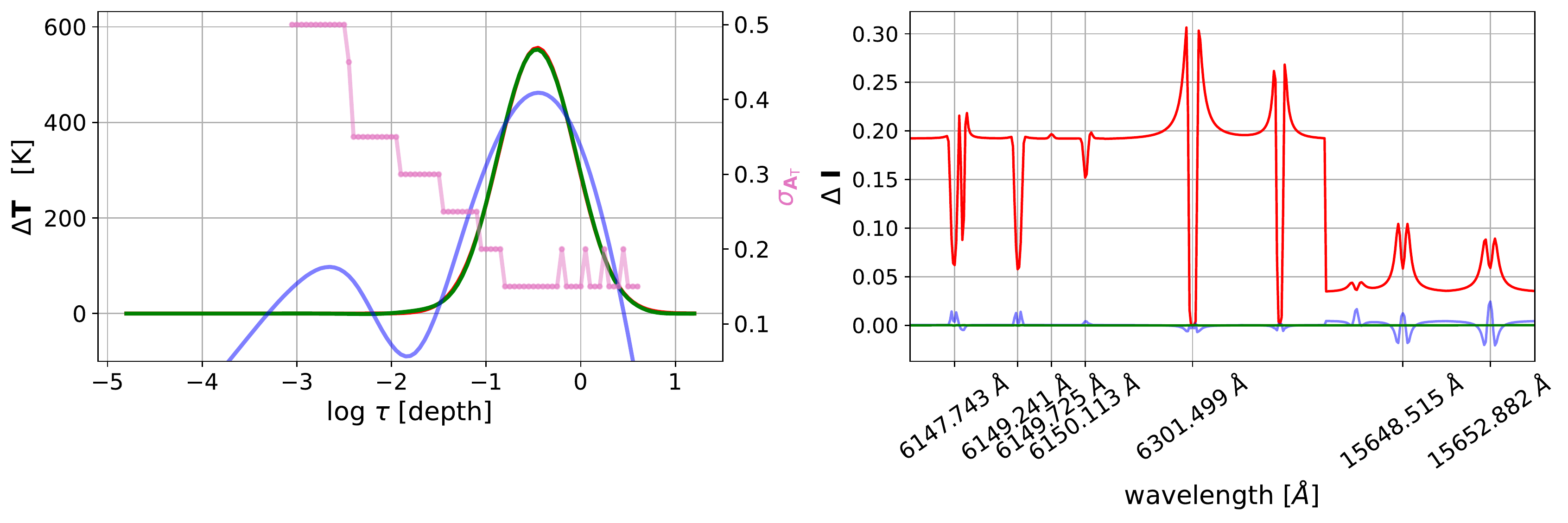}			
			    \end{center} 
			    \caption{{Lefthand plot}: Temperature perturbation to be inverted, replotted from the top-left panel in Figure~\ref{fig4_1_testcase1_model_and_spectra} in red. Iterative OLA inversion result is shown in green, that obtained using SIR is shown in blue (using the typical SIR parameters, as stated in Appendix~\ref{appendix_sir}). The minimum kernel widths achievable in the final iteration cycle which also approximate the vertical resolution of OLA solutions, are shown in pink (plotted with a different scaling). Righthand plot:  The corresponding spectral differences \dSIv\ for the starting guess model spectra (red), OLA inverted model spectra (green), and SIR inverted model spectra (blue). Note that the L1 norm of observed $-$ guess \SIv\ is 57.4, while those for iterative OLA inverted model and SIR inverted model are 0.03 and 0.947, respectively. \\ }	
			    \label{fig4_3f:olainvtest1-final-inversion-plot}
			\end{figure*}
		
		While this first inversion cycle solution approximates a running average of the underlying perturbation, it is irregular in shape and has sharp edges at the OLA inversion window boundary. Unlike SIR, OLA doesn't aim to obtain a globally smooth solution, and since the inversions are computed independently at each depth, at each depth they feel different \coeffv $^\top \bigcdot$ \errv\ and cross-talk error contributions. 
		Using the outcome of the iteration directly in the next inversion cycle iteration proves problematic.  Since the starting model for the next iteration is obtained by updating the guess \Tv\ profile with result from the current iteration cycle, it will reflect the irregularities of the solution, and importantly, since the response functions must be recalculated with each new iteration, they too will not be smooth.  
		This makes it hard to construct, in subsequent cycles, averaging kernels that match smooth target function profiles. Thus, we employ a smoothed version of the inverted \dTv\ in the next iteration.  The smoothing is done via convolution with a Gaussian kernel of width equal to that of the narrowest averaging kernel achieved in the previous inversion cycle. Note that smoothing smears the inverted solution outside of the inversion window which helps with the 'edge-effect' issue (see Section~\ref{subsec4_4:largescale-dT}).  The result for the first inversion cycle is plotted as the green-solid curve in the left panel of Figure~\ref{fig4_5_test1_lowhigh_rank_soln}.  It is this solution that is added to initial guess to obtain the inverted \Tv. 

		Before moving on to the next iteration cycle, a decision is required on whether to continue inverting using the same \pinv\ rank, increase the rank, or stop the iteration altogether. This decision is based on a comparison between the L1 norm of \dSIv\ calculated from the  inverted atmosphere and that of  \dSIv\ obtained with the guess model.  If the spectral fit improved, the next inversion cycle employs the same \pinv\ rank.
		When using the same rank in successive iterations, the magnitude of iterative updates get smaller, along with the improvement in the spectral fit. The magnitude of \fdTvflat\ decreases with iteration.  Additionally, with constant $k$ iteration, \fdTvflat\ becomes more oscillatory, 
		as each successive cycle refines the previous approximation.   
			Moreover, while the response functions are recalculated in each cycle, the kernel widths that are achievable at each depth do not change significantly when the same rank \pinv\ is employed.  They are unable to resolve the oscillatory residual, and a larger rank is required to construct narrower kernels and make further improvements. We determine that the iterative updates at fixed rank have 'stagnated' if the difference between the L1 norm of guess \dSIv\ and that of inverted model \dSIv\ (in a cycle) are within $0.5\%$ of one another. 
			When this is the case, we increase rank by two (an iteration acceleration arrived at by trial-and-error) and proceed with the next cycle. 
			Inversion iteration thus proceeds in successive stages of rank increase and stagnancy, until the rank can not be further increase without error amplification.  			
		
		In Figure~\ref{fig4_3f:olainvtest1-final-inversion-plot} (left plot), the green curve shows the inversion result after employing iterative OLA method (barely distinguishable from the underlying perturbation shown in red, which it overlies).  A SIR inversion for the same perturbation is plotted in blue. While iterative OLA recovers the underlying perturbation well, SIR struggles to recover it.  This is largely because SIR, by design, inverts at all depths.  This includes regions well outside of the OLA inversion window where the solution is poorly constrained by the response functions and can be dominated by contributions from other depths.  While poorly constrained, these regions none-the-less contribute to the global spectrum and, as illustrated by the right-hand panel of Figure~\ref{fig4_3f:olainvtest1-final-inversion-plot}.  Both the SIR and iterative OLA inversions yield comparably good fits to the observed spectrum.

		To ensure that the final inverted solution is close to the underlying atmosphere, it is critical to minimize the both cross-talk error and contributions from \errv, especially early in the iterative process when \dSIv\ is the largest.  With successive updates \dSIv\ gets smaller and it gets increasingly harder to make significant changes to solution already at hand. The SIR solution allows more cross-talk error from regions that are poorly constrained.  The iterative OLA approach, on the other hand,  can optimally minimize cross-talk contributions, but struggles when non-zero perturbations exist outside the reliably invertible region, regions that are not updated during the inversion iteration. When iteration is required, the limiting inversion window introduces an 'edge-effect.'  This is the subject of the next section. 

	\subsection{Iterative OLA method: the "edge-effect" issue} \label{subsec4_4:largescale-dT}	
		\begin{figure*}[!t] 
			\begin{center}
			\includegraphics[width=.98\textwidth]{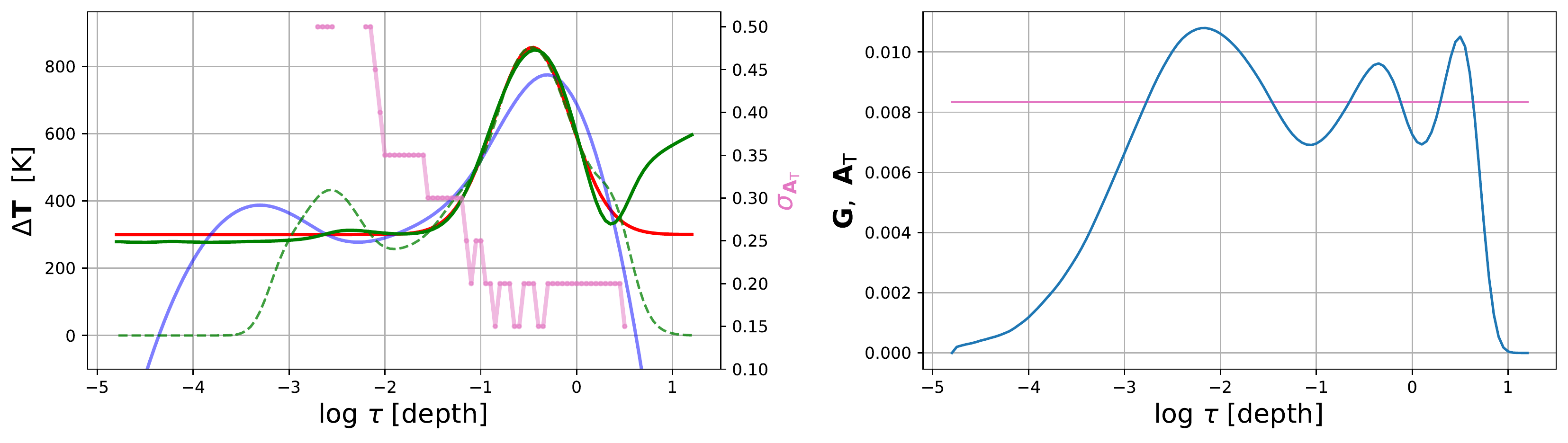}
		    \end{center}
		    \caption{{Lefthand plot}: Inversion results for Gaussian shaped localized perturbation (same as red curve in the left panel in Figure~\ref{fig4_3f:olainvtest1-final-inversion-plot}) combined with a constant 300 K offset (in red).  Inverse solution obtained using iterative OLA method is shown in green-dashed and that obtained using iterative OLA combined with flat large-scale kernel updates (Section~\ref{subsec4_4:largescale-dT}) is shown in green-solid.  Solution obtained using SIR is shown in blue. Pink curve corresponds to the minimum kernel width of the final localized Gaussian kernel when employing it along with the flat kernel large-scale updates. 
		    {Righthand plot}:  The flat target function (pink, area normalized to 1) employed to capture the mean offset, and the large-scale averaging kernel achieved (blue), when constructed using rank $k=5$ \pinv\ of the mean \muram\ \fRFTm. \\ } 
		    \label{fig_large_scale_inv}		
		\end{figure*}

	The Gaussian-shaped perturbation employed in the last section to illustrate the application of the OLA method (left plot in Figure~\ref{fig4_3f:olainvtest1-final-inversion-plot}), while of sufficiently large amplitude to require an iterative solution, was placed in a region of the domain where averaging kernels could readily be constructed.  This favored a successful OLA inversion. Other perturbations present greater challenge.  Somewhat surprisingly, a simple large-scale offset between the underlying and guess models is particularly difficult for OLA to uncover. 
	An example is shown in red in the left panel in Figure~\ref{fig_large_scale_inv}, where in addition to the localized Gaussian of last section, a constant offset of 300 K is added to the atmosphere.  In this case, the iterative OLA method struggles to recover the perturbation (inversion result shown with a green-dashed curve). 	

	The failure of the OLA method in this case has two underlying and intertwined causes:  iterative updates are confined to a limited depth range outside of which updates are not made (localized kernels can't be constructed in those regions), and the averaging kernels have finite widths, they are not $\delta-$functions. With successive updates, the underlying \fdTvflat\ within the inversion window gets smaller but outside of that window it does not change much {(smoothing inverted \dTv\ at the end of an iteration cycle to remove irregularities in the solution smears them outside the window)}.  
		Thus, with iteration, the averaging kernels on which the inversion solutions are based on, if they admit contributions from outside of the inversion window, increasingly weigh the uncorrected perturbations with iteration. There is consequently leakage of information from the outside of the inversion window into it (the solution becomes cross-talk error dominated), which degrades the overall inversion quality.  While smoothing inversion solutions, increasing the rank and allowing the inversion window to grow with iteration help and were shown in the example illustrated by the green-dashed curve in Figure~\ref{fig_large_scale_inv}, updates cannot typically be made at all depths and the problem can not be entirely avoided. 
	
	One solution is to extend the inversion updates to outside of the OLA inversion window, and thus to minimize its negative influence.  This can be done by including an averaging kernel that captures the large scale perturbations that extend into those regions.  For example, to capture a constant offset, one can define a flat (constant) normalized target functions (pink curve in the right panel in Figure~\ref{fig_large_scale_inv}) and use the OLA framework to construct flat averaging kernels that are equally sensitive at all depth locations, and interleave inversions based on this kernel with the those using localized Gaussian kernels during iteration.  If such a flat kernel can be constructed, an inversion solution based on it \coeffvt $\bigcdot$ \dSIv\ would correspond to the large-scale average fractional contribution, \avg{\fdTv} $= \mathbf{A}_\text{T}^\top \bigcdot$ \fdTv $\ \equiv \frac{1}{N_\tau}\sum\limits_{\tau}$ \fdTi{}, where $N_\tau$ is the total number of optical depth points. In practice, the flat averaging kernel only approximates a constant with optical depth (blue curve in the right panel in Figure~\ref{fig_large_scale_inv}). The average is also an approximation.   

	The final inversion solution from this combined iterative effort is shown with the green-solid curve in the left panel of Figure~\ref{fig_large_scale_inv}. While the solution is improved, some difficulties remain.  Since the large-scale inversion is for the fractional average \avg{\fdTv} (from which  \dTi{i} at each depth is computed via \dTi{i}  = \avg{\fdTv} $\times \ \text{guess T}(\tau_i)$) and since T($\tau_i$) in the deeper regions is about 3-5 times larger than those in the shallower region, the method tends to overcorrect in the deeper regions and under-correct in the shallower regions. 
	Employing non-fractional response functions and directly computing constant inverted \dTv\ (and not \fdTvflat) yields, in this case, a more robust solution, but induces similar issues for cases in which underlying perturbation is not a constant offset.
	 Given that we do not know a priori the behavior of the underlying large-scale perturbation, it is hard to determine the target function that best captures it. 
	 
	Another possible solution to the edge-effect problem, is to employ non-symmetrical target functions that have low amplitudes outside of the inversion window. Unfortunately, just as the response functions typically available do not allow $\delta-$function averaging-kernel construction, they do not allow construction of an infinitely sharp cutoff. Since the magnitude of the perturbation inside the inversion window decreases with iteration, the importance of any residual kernel amplitude outside the inversion window increases if the underlying targeted perturbation has any magnitude in that region.   
	
	Finally, since the nodal approach of SIR is adept at recovering large-scale constant or linear trends, we developed a hybrid SIR/iterative OLA scheme.  This approach met with some success (see Section~4 of \cite{ourthesis}), and is the subject of ongoing research efforts.

\section{Multivariable inversion using iterative OLA method} \label{sec_multivar}
	\begin{figure}[!t] 		
			\includegraphics[width=0.47\textwidth]{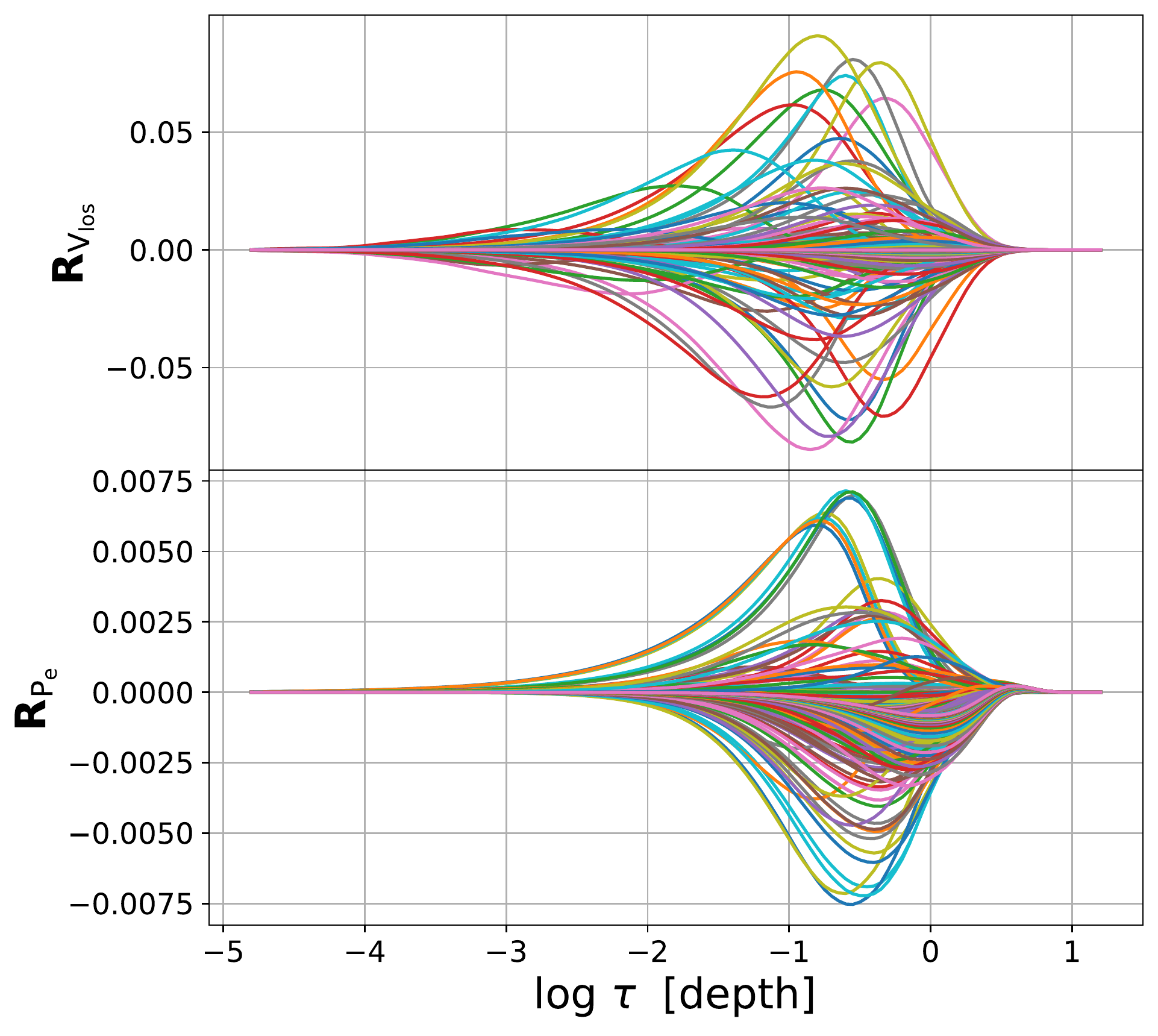}
		      \caption{({Top panel}) Fractional line-of-sight velocity response functions as a function of optical depth for individual wavelengths that make up the observed spectrum. Note, adiabatic sound speed $\mathbf{c_s}$ is used to non-dimensionalize the response functions, as terms in \Vlosv\ can be 0. ({Bottom panel}) Fractional response functions for electronic pressure. Both sets of response functions were computed using SIR for mean \muram\ model. \\}
		    \label{fig_RFPe_RFVlos}			
		\end{figure}

	The OLA method can be extended to include spectral sensitivity to more than one variable.  Here, we examine multivariable inversions for the thermodynamic parameters temperature, electronic pressure, and line-of-sight velocity, \Tv, \Pev\ and \Vlosv, under the simplifying assumption that these completely define the atmosphere.

	The 1st order equation for a multivariable system can be written as  
		\begin{eqnarray} \label{eqn6_1:multivar1storder}
			\begin{aligned}
			\Delta\mathbf{I} \  &= \  \mathbf{R}_\text{T}^\top \bigcdot \frac{\Delta \mathbf{T}}{\mathbf{T}} \ + \ \mathbf{R}_{\text{V}_\text{los}}^\top \bigcdot \frac{\Delta \mathbf{V_{los}}}{\mathbf{c_{s}}} \ \\ & \ \ \
			\  + \  \mathbf{R}_{\text{P}_\text{e}}^\top \bigcdot \frac{\Delta \mathbf{P_e}}{\mathbf{P_e}} \ + \ \boldsymbol{\varepsilon} , 
			\end{aligned}
		\end{eqnarray}

		\noindent
		where, \fRFVlosm\ and \fRFPem\ represent the response function matrices for \Vlosv\ and \Pev\  (Figure~\ref{fig_RFPe_RFVlos}), and $\mathbf{R}_\text{T}$ is that for \Tv, as previously.  Note the use of adiabatic sound speed $\mathbf{c_s}$ when computing the fractional response function \fRFVlosm\ to avoid difficulties where \Vlosv\ is close to zero. Further note, that \errv\ now collectively represents the missing higher order error terms from all variables, plus observational noise (if present). 
		
	\subsection{Averaging and cross-talk kernels} \label{chp6_multivar_akconst}

		The underlying strategy for multivariate OLA inversions is similar to the single variable counterpart; individually invert for each variable at each depth location, with the goal of computing solutions that optimally balance \errv\ and cross-talk errors. In multivariate inversions, there are cross-talk error contributions from other depths, as previously in single variable inversions, and from other variables. 
		The optimally error balanced solution is achieved by computing a set of coefficients \coeffv\  that can be used to linearly combine the response functions of the target variable to form a narrow Gaussian averaging kernel, while at the same time (using the same coefficients), can be used to combine the response functions for the cross-talk variables so that their sum is nearly zero at all depths. 
		For example, if the target is to invert \Tv\ at depth $\tau_i$, and the cross-talk variables are \Vlosv\ and \Pev, then the coefficients \coeffv\  are determined by solving the linear system of equations, 
				\begin{eqnarray} \label{eqn_akconst_multivar} 
					\begin{pmatrix}
						\mathbf{R}_\text{T} \\ \mathbf{R}_{\text{V}_{\text{los}}} \\ \mathbf{R}_{\text{P}_{\text{e}}} 
					\end{pmatrix} 
					\bigcdot \mathbf{C} \ &=& \ 
					\begin{pmatrix}
						\mathbf{G}\ (\tau_i, \sigma) \\ 0 \\ 0
					\end{pmatrix} .
				\end{eqnarray}	

		\noindent
		The coefficients \coeffv\ yield the averaging kernel  \AKTvi{i} = \fRFTm\ $\bigcdot$ \coeffv\  which approximates $\mathbf{G} (\tau_i, \sigma)$, while minimizing cross-term sensitivities with averaging kernels \ctAKVlosv\ = \fRFVlosm $\bigcdot$ \coeffv\ and \ctAKPev\   = \fRFPem $\bigcdot$ \coeffv\ which are approximately zero. 
		The kernels \ctAKVlosv\ and \ctAKPev\ capture the $\tau$ dependence of the residual cross-talk. 
		Once coefficients are computed, the dot product of \coeffv$^\top$ with Equation~\ref{eqn6_1:multivar1storder} yields the inverse solution

				\begin{eqnarray} \label{eqn6_3:multivarsoln} 
					\begin{aligned}
						\mathbf{C^\top} \ \bigcdot \ \Delta\mathbf{I} \ &= \ \mathbf{A}_\text{T}^\top (\tau_i) \bigcdot \frac{\Delta \mathbf{T}}{\mathbf{T}} \ + \ \mathbf{A}_{\text{V}_\text{los}}^\top (\tau_i) \bigcdot \frac{\Delta \mathbf{V_{los}}}{\mathbf{c_{s}}} \ \\ & \ \ \ \ + \ \mathbf{A}_{\text{P}_\text{e}}^\top (\tau_i) \bigcdot \frac{\Delta \mathbf{P_e}}{\mathbf{P_e}} \ + \ \mathbf{C^\top} \bigcdot \boldsymbol{\varepsilon} .
					\end{aligned}
				\end{eqnarray}
				
		\noindent
		 If Equation~\ref{eqn_akconst_multivar} had an exact solution, \ctAKVlosv\ and \ctAKPev\ would be exactly zero at all depths, and  
		 \coeffv $^\top \bigcdot$ \dSIv\ would correspond to the \AKTv\ weighted average, \avg{\fdTi{i}} (assuming also negligible error contribution \coeffv $^\top \bigcdot$ \errv).  Importantly, such an exact solution would not only recover \avg{\fdTi{i}} uncontaminated by cross-talk variable contributions, but the spectral contributions of the cross-talk variables to \dSIv\ (Equation~\ref{eqn6_1:multivar1storder}) would be zero and would thus be preserved for cross-talk variable inversion.

		However, Equation~\ref{eqn_akconst_multivar} can only be approximately solved.   The solution based on a lower rank \pinv\ of the response function matrix does not completely suppress \ctAKVlosv\ and \ctAKPev, and the contributions of $\Delta \mathbf{V_{los}}/{\mathbf{c_{s}}}$ and ${\Delta \mathbf{P_e}}/{\mathbf{P_e}}$  to the inversion solution are additional error sources. Further, in a multivariable system, there is a trade-off between the minimum averaging kernel width that can be constructed for the target variable and the suppression of cross-talk sensitivity to other variables.  This trade-off is influenced by the underlying spectral-sensitivity differences between the variables to atmospheric properties.

	\subsection{Variable spectral sensitivity bias and its correction}
	\label{chp6_multvar1_rfamp}
			\begin{figure}[!t] 		
			    \begin{center}
					\includegraphics[width=0.475\textwidth]{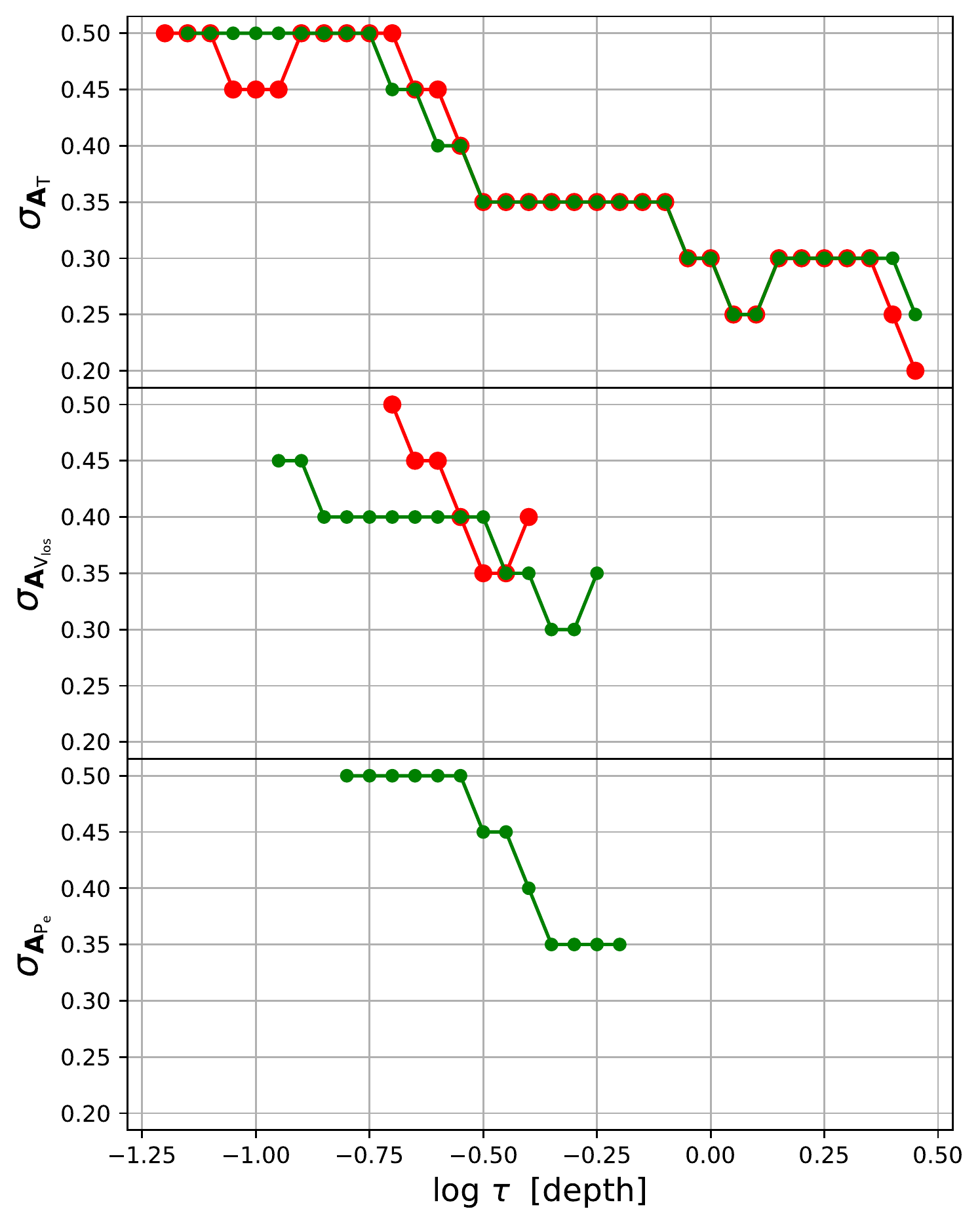}
				\end{center} 
			    \caption{({Top panel}) Minimum temperature kernel widths achievable at each depth, in a multivariable system, with (green) and without (red) response function amplification factors (see Section~\ref{chp6_multvar1_rfamp}). Similarly, {middle} and {bottom} panels corresponds to minimum widths achievable at each depth for line-of-sight velocity and electronic pressure. Note, in all cases, \pinv\ matrix is computed using rank $k$ such that the dominant singular values add up to $95\%$ of the total sum. This corresponds to $k=10$ (with) and $k=7$ (without) amplification factors. \\ }
			    \label{fig_multivar_GTFwid_vs_depth}				
			\end{figure}

			\begin{figure*}[!t] 		
			    \begin{center}
					\includegraphics[width=.98\textwidth]{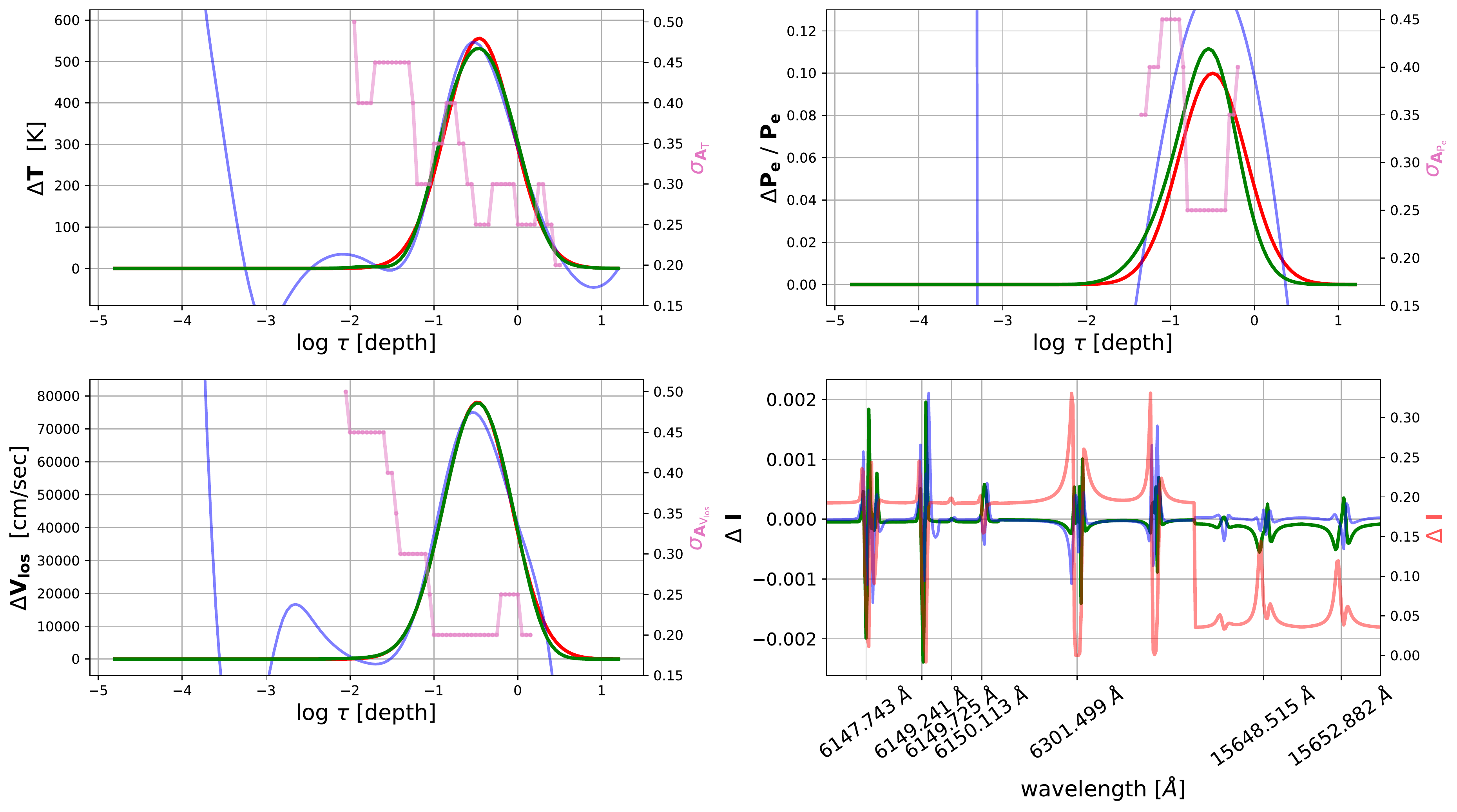}
			    \end{center} 
			    \caption{Multivariable inversion of localized perturbations are shown here. 
			    Red curve in each panel (except bottom-right) corresponds to the Gaussian perturbation (peak location at \logtau\ $= -0.5, \text{width} = 0.40\ \Delta$ \logtau\ and perturbation amplitude = $10\%$) added to the corresponding variable in the mean \muram\ model, to construct underlying model. The final solution obtained using iterative OLA method (without large-scale flat kernel updates) and SIR are shown in green and blue, respectively.
			    Pink curve, plotted with a different scaling, corresponds to the minimum kernel widths achievable for each variable in the final OLA iteration cycle. 
			    {Bottom-right} panel shows corresponding spectral differences \dSIv\ for the OLA inverted model spectra (green) and SIR inverted model spectra (blue), and starting guess model spectra (red, plotted with a different scaling). \\}
			    \label{fig_multivar_iter_OLA}
			\end{figure*}

		Spectral sensitivity varies with atmospheric variable.  Typically, the spectral response to temperature perturbations is greater than that due to electronic pressure or \los\ velocity perturbations. This is reflected in the magnitudes of the fractional response functions 
		(compare the magnitudes of the response functions making up \fRFTm\ in Figure~\ref{fig4_1_testcase1_model_and_spectra} to those of \fRFVlosm, \fRFPem\ in Figure~\ref{fig_RFPe_RFVlos}). The peak amplitude of \fRFTm\ is about 30 times larger than that of \fRFPem and about 3 times larger than that of \fRFVlosm. 
		Thus the spectral sensitivity (in Equation~\ref{eqn6_1:multivar1storder}) and the determination of the kernel coefficients (e.g. in Equation~\ref{eqn_akconst_multivar}) are biased toward temperature.  The OLA coefficients satisfy \fRFTm\ $\bigcdot$ \coeffv\ = \GTFv\ more closely than \fRFPem\ $\bigcdot$ \coeffv\ = 0 or \fRFVlosm\ $\bigcdot$ \coeffv\ = 0, making it is easier to construct  \AKTv\ than minimize \AKPev\ or \AKVlosv. This is true also when the target is a variable other than \Tv, in which case the zero amplitude kernel \AKTv\ for the suppression of \Tv\ contributions is favored over the Gaussian kernel for the target variable.   
		
		This is illustrated by {red} curves in Figure~\ref{fig_multivar_GTFwid_vs_depth} which show that, for a given rank \pinv\ matrix, narrower kernels can be constructed in the multivariable system (i.e., while suppressing the non-target variables with zero amplitude target functions) over a wider depth range for temperature (top plot) than they can for \los\ velocity (middle plot) or electronic pressure (bottom plot). The variation in response function sensitivities leads to the larger singular values being associated with the modes that capture \fRFTm\  sensitivities, while those which capture \fRFPem\ and \fRFVlosm\ correspond to smaller singular values. When a lower rank \pinv\ matrix is employed to prevent \errv\ dominated solutions, the \fRFPem\ and \fRFVlosm\ sensitivities are truncated more severely than \fRFTm, and the computed coefficients are biased towards satisfying the temperature related portions of the linear system.    
		
		The biased sensitivity of the reduced rank \pinv\ matrix is particularly critical when constructing \AKPev\ for a \fdPevflat\ inversion which requires the simultaneous suppression of the \fdTvflat\ and \fdVlosvflat\ contributions. In the solution of
			\begin{eqnarray} \label{eqn_akconst_multivarPe} 
				\begin{pmatrix}
					\mathbf{R}_\text{T} \\ \mathbf{R}_{\text{V}_{\text{los}}} \\ \mathbf{R}_{\text{P}_{\text{e}}} 
				\end{pmatrix} 
				\bigcdot \mathbf{C} \ &=& \ 
				\begin{pmatrix}
					0 \\ 0 \\ \mathbf{G} (\tau_i, \sigma) 
				\end{pmatrix} ,
			\end{eqnarray}

		\noindent
		the smaller \fRFPem\ sensitivity prevents construction of even a very wide averaging kernel at any depth, when employing the 95\% criterion on the \pinv\ rank (no {red} curve in the lower panel of Figure~\ref{fig_multivar_GTFwid_vs_depth}).   The coefficients computed  largely ensure that \fRFTm\ $\bigcdot$ \coeffv\ = 0, with little sensitivity to  \fRFPem\ $\bigcdot$ \coeffv\ =  \GTFv. The constructed \AKPev\ do not fit any allowed width target function (at any depth) with at least $80\%$ accuracy (see Section~\ref{sec_olainv_ata_depth}). While it is possible to construct \AKPev\ by employing larger rank, computed solutions are then more likely to become \errv\ dominated.	

		One way to address the differential spectral sensitivity is to artificially amplify/de-amplify the spectral response functions so that sensitivities to all variables are similar. For example, in place of Equation~\ref{eqn_akconst_multivarPe}, a modified system of equations
				\begin{eqnarray} \label{eqn_akconst_multivar2Pe} 
					\begin{pmatrix}
						\alpha_1 \cdot \mathbf{R}_\text{T} \\ \alpha_2 \cdot \mathbf{R}_{\text{V}_{\text{los}}} \\ \mathbf{R}_{\text{P}_{\text{e}}} 
					\end{pmatrix} 
					\bigcdot \mathbf{C} \ &=& \ 
					\begin{pmatrix}
						0 \\ 0 \\ \mathbf{G} (\tau_i, \sigma) 
					\end{pmatrix} 
				\end{eqnarray}

		\noindent
		is solved, where $\alpha_1 \ \text{and} \ \alpha_2$ are the response function amplification factors used to amplify/de-amplify cross-talk variable sensitivities.  While this mathematical operation is equivalent to multiplying both sides of the equation by a constant factor and thus does not change the meaning of the original equation, it does change the singular values and \pinv\ matrix employed, manually forcing the underlying eigenvectors to carry similar sensitivities across all variables. 
		
		It is critical to choose the amplification factor judiciously. If too large, the lower rank \pinv\ matrix carries enhanced cross-talk variable sensitivity resulting in a solution that primarily minimize cross-talk at the expense of target variable kernel construction, and narrowing the averaging kernel width achievable remains difficult. 
		If too small, the lower rank \pinv\ matrix carries insufficient cross-talk sensitivity, and cross-talk is not actively suppressed.
		Empirically, we have found that the ratio of the maximum of the absolute value of response functions works well i.e. in Equation~\ref{eqn_akconst_multivar2Pe}, $\alpha_1 =  \frac{max(\lvert \mathbf{R}_{\text{P}_\text{e}} \rvert)}{max(\lvert \mathbf{R}_{\text{T}} \rvert)}$ and $\alpha_2 = \frac{max (\lvert \mathbf{R}_{\text{P}_\text{e}} \rvert)}{max (\lvert \mathbf{R}_{\text{V}_\text{los}} \rvert)}$. For the response function matrices obtained from the mean \muram\ model, these ratios are $\alpha_1 \approx 0.03\ \text{and}\ \alpha_2 \approx 0.08$. These factors are less than 1 because the magnitude of terms in \fRFTm\ and \fRFVlosm\ is larger than \fRFPem.   We have chosen, independent of the target variable, to apply the amplification factors to the cross-talk variable response functions.  This avoids the need to factor out the amplification factors from the coefficients before computing the inverse solution \coeffv\ $^\top \bigcdot$ \dSIv. 

		When the amplification factors are employed, the dominant modes of the \pinv\ matrices retain sensitivity to the variables with intrinsically lower amplitude spectral response functions.  This is shown with the green curves in Figure~\ref{fig_multivar_GTFwid_vs_depth}.  In the particular case of \fdPevflat\ inversions, averaging kernels \AKPev\ can be constructed in the presence of both T and \Vlos\ maintaining that 95\% criterion determine the \pinv\ rank. Notably, a broader range of depths become accessible to \Vlos\ inversions, and \AKPev\ construction and \fdPevflat\ inversion becomes possible (bottom panel). 
		
		\subsection{Multivariable Iterative OLA inversion: Example Implementation}

			We have carried out multivariable iterative OLA inversions (without large-scale offset correction) for idealized perturbations of the mean \muram\ atmosphere.  Equations like that of Equation~\ref{eqn_akconst_multivar2Pe} were used to invert for each variable at each depth, employing a minimum width Gaussian target function for the target variable while simultaneously preserving the spectral contributions of the cross-talk variable via null target functions, as described above.  OLA, by design, inverts for only one variable at a time, and thus each inversion cycle includes successive iteration over each variable. The order we have chosen to invert for each variables (at all depths for which a kernel can be constructed) is in decreasing order of their spectral sensitivity, first for \fdTvflat, then for \fdVlosvflat, and subsequently for \fdPevflat. This order was chosen because it is easier to construct narrow kernels for the spectrally more influential variables and suppress sensitivities of the less dominant ones than vice versa.  
			
			In implementation, within a given cycle, we update the guess \Tv\ with inverted \Tv, if inverted \fdTvflat\ results in a better spectral fit and then repeat the inversion process for the next dominant variable \fdVlosvflat, and finally \fdPevflat, each with the same spectral fit criterion. 
			Thus the final inverted model (after any full cycle) is updated only with those variables that help improve the spectral fit measure and only at those depths at which a kernel can be constructed (combined with smoothing that smears inversion updates outside the inversion window). This approach helps reduce the corresponding perturbation magnitudes, which lessens overall \errv\ magnitude. A smaller \errv\ implies larger rank $k$ can be employed in subsequent cycles, improving the OLA inversion resolution and depth range, and thus the inversion quality achievable.  Similar to its single variable counterpart, we compare guess spectral fit with that obtained from all variable updated inverted model to determine what rank to use for the next cycle or to stop iteration altogether. 

			In Figure~\ref{fig_multivar_iter_OLA} we show a test case perturbation (red) obtained by adding Gaussian perturbations to the mean \muram\ profile of each variable (peak location at \logtau\ $= -0.5$, width = $0.40\ \Delta$ \logtau), with perturbation amplitudes equal to $10\%$ of the variable value at the peak location depth. The final inverted solution obtained using iterative OLA  and SIR are shown in green and blue. OLA and SIR are both able to recover the Gaussian perturbations. Iterative OLA localizes it quite robustly, while the SIR solution shows strong oscillatory wings in deeper and shallower regions.  It is important to note however, that success by iterative OLA solver is favored by this particular case.   The underlying perturbation we have chosen for illustration does not have a large-scale offset and thus the edge-effect issue discussed in Section~\ref{subsec4_4:largescale-dT} has been avoided.  SIR inversions are much more successful at recovering such large-scale offsets between the model and observed atmosphere, and we are pursuing, among other possibilities, a hybrid approach that can robustly capture both large-scale and localized perturbations and yield quantitative assessment of the depth resolution achieved.

\section{Summary and future work} 
\label{sec_conclusion}

		In this work, we applied the OLA method to synthetic spectra to invert for the thermodynamic properties (\Tv, \Pev\ and \Vlosv) of a simulated solar atmosphere, assuming that the atmosphere and radiation field are in LTE.  These inversions, as previous, solve a 1st order linear system of equations for difference between the guess model and that for which the spectrum is observed.  The spectral sensitivities are captured by the spectral response functions and an iterative scheme is employed to relax the model to a state in which the observed spectrum and that from inverted model agree.  In our case, as the observed atmosphere is synthetic, the final atmospheric properties can be compared for agreement. 
		
		Inverted solutions have two intrinsic error contributions: error \errv\ from the higher order terms truncated by the linear response function formalism plus observational/instrumental noise, if present, and error from cross-talk, the erroneous contributions from other depths and other variables in a multi-depth multivariable solution. It is generally not possible to minimize cross-talk error without amplifying \errv, and vice versa. The goal of inversions is to regularize and balance these error contributions by employing a truncated rank \pinv\ matrix. Employing a smaller rank results in cross-talk error dominated solutions, while larger rank solutions are dominated by \errv. Additionally, due to the non-local  nature of the radiative transfer, different atmospheric solutions are consistent with a given spectral difference. This makes it hard to determine the rank that optimally balances the two errors.  
		Current state-of-the-art spectral-inversion methods such as SIR, compute globally smooth solutions that minimizes \errv, but these solutions are likely cross-talk error dominated.  It is hard for these approaches to recover steep gradients, or a priori know what gradients are recoverable given the spectral response function set on hand.  Addressing these issues is the key motive behind OLA. 

		OLA independently inverts for each variable at each depth location, to obtain solutions that optimally balances \errv\ and cross-talk contributions. This is achieved by solving for the coefficients which can be used to construct, as a linear combination of the response functions, a localized averaging kernels that mimics prescribed target functions, those that both localize depth sensitivity to the target variable and zero-out sensitivity to other variables. The inverted solution at a given depth, given by the inner product of the coefficients with the spectral difference approximates the kernel averaged underlying field. The width of the kernel thus represents the spatial resolution of the inverted solution. The process is repeated to invert at all possible depths. Given the limited orthogonal sensitivity of the response functions, localized kernels cannot be constructed at all depths. Failure usually occurs above and below a limit range of optical depths, defining the 'OLA inversion window.'  The rank of the \pinv\ matrix dictates the size of this window, the kernel width achievable at each depth, and the error regime the inverted solution  falls into. We developed an iterative OLA that relaxes the solution, with increasing rank with iteration, from the likely non-linear differences between the initial model and the observed atmosphere to the final model which matches the spectral observations.

		We have applied the iterative OLA method we have developed to inversions for large amplitude perturbations using linear response functions that are iteratively updated, significantly extending the inversion capability of traditional OLA methods used in helioseismology.  The inversion results we achieve are promising and competitive with other state-of-the-art spectral inversion techniques.  None-the-less, some issues remain. Iterative OLA struggles to recover large-scale offsets due to non-zero perturbations outside the OLA inversion window.  This is a direct consequence of the inability to construct strictly localized $\delta$-function kernels. Iteratively updating regions inside the inversion window eventually causes a leakage of the uncorrected perturbation information from the outside of the inversion window to the inside of it.  In the worst cases this can corrupt the entire solution, and in doing so defeats the very strength of the OLA method, which fundamentally aims to minimize leakage by minimizing cross-talk and thus preserve the locality of the underlying perturbations. 

		We have shown, that the edge-effect issue can be minimized by making updates outside of the OLA inversion window. For this, we developed a scheme within the OLA framework where we construct 'flat' averaging kernels (instead of Gaussian-shaped localized kernels) to obtain a large-scale averaged solution. 
		We interleaved large-scale averaged solution with the high-resolution localized OLA inversion during iteration. While the inversion quality is significantly improved, some difficulties remain, and finding a more robust solution this problem is a future endeavor. 

		We have also shown that the inherent spectral sensitivity bias in multivariable inversions can be overcome using response function amplification, which allows dominant modes of the \pinv\ matrix to be equally sensitive to all variable. We expect that this idea would benefit other inversion methods as well, and have shown that its implementation can allow for inversions of electronic pressure, which is typically difficult to achieve.  
		
		This work was partially supported by National Science Foundation grant number 1616538.  Additionally, PA acknowledges the support of the University of Colorado's George Ellery Hale Graduate Student Fellowship, and MPR recognizes a College Scholar Award from the University of Colorado, Boulder, which was made possible through the generosity of donors to the College of Arts and Sciences.  PA would like to thank Ivan Milic, Han Uitenbroek, Gianna Cauzzi, and Kevin Reardon for their help and support.  The research made use of NASA's Astrophysics Data System Bibliographic Services.

\bibliography{refs.bib}{}

\begin{thebibliography}{}
\expandafter\ifx\csname natexlab\endcsname\relax\def\natexlab#1{#1}\fi
\providecommand{\url}[1]{\href{#1}{#1}}
\providecommand{\dodoi}[1]{doi:~\href{http://doi.org/#1}{\nolinkurl{#1}}}
\providecommand{\doeprint}[1]{\href{http://ascl.net/#1}{\nolinkurl{http://ascl.net/#1}}}
\providecommand{\doarXiv}[1]{\href{https://arxiv.org/abs/#1}{\nolinkurl{https://arxiv.org/abs/#1}}}

\bibitem[{{Agrawal}(2021)}]{ourthesis}
{Agrawal}, P. 2021, PhD thesis, University of Colorado at Boulder.
\newblock \url{https://ui.adsabs.harvard.edu/abs/2021PhDT.........7A}

\bibitem[{Anan {et~al.}(2021)Anan, Schad, Kitai, Dima, Jaeggli, Tarr, Collados,
  Dominguez-Tagle, \& Kleint}]{anan21}
Anan, T., Schad, T.~A., Kitai, R., {et~al.} 2021, The Astrophysical Journal,
  921, 39, \dodoi{10.3847/1538-4357/ac1b9c}

\bibitem[{{Asensio Ramos} {et~al.}(2008){Asensio Ramos}, {Trujillo Bueno}, \&
  {Landi Degl'Innocenti}}]{ramos08}
{Asensio Ramos}, A., {Trujillo Bueno}, J., \& {Landi Degl'Innocenti}, E. 2008,
  \apj, 683, 542, \dodoi{10.1086/589433}

\bibitem[{{Auer} {et~al.}(1977){Auer}, {Heasley}, \& {House}}]{auer77}
{Auer}, L.~H., {Heasley}, J.~N., \& {House}, L.~L. 1977, \solphys, 55, 47,
  \dodoi{10.1007/BF00150873}

\bibitem[{{Backus} \& {Gilbert}(1968)}]{backus68}
{Backus}, G., \& {Gilbert}, F. 1968, Geophysical Journal, 16, 169,
  \dodoi{10.1111/j.1365-246X.1968.tb00216.x}

\bibitem[{{Backus} \& {Gilbert}(1970)}]{backus70}
---. 1970, Philosophical Transactions of the Royal Society of London Series A,
  266, 123, \dodoi{10.1098/rsta.1970.0005}

\bibitem[{Backus \& Gilbert(1967)}]{backus67}
Backus, G.~E., \& Gilbert, J. 1967, Geophysical Journal International, 13, 247

\bibitem[{Basu(2016)}]{basu2016}
Basu, S. 2016, Living Reviews in Solar Physics, 13, 2,
  \dodoi{10.1007/s41116-016-0003-4}

\bibitem[{Bellot~Rubio \& Orozco~Su{\'a}rez(2019)}]{rubio19}
Bellot~Rubio, L., \& Orozco~Su{\'a}rez, D. 2019, Living Reviews in Solar
  Physics, 16, 1

\bibitem[{{Bellot Rubio}(2006)}]{luis06}
{Bellot Rubio}, L.~R. 2006, in Astronomical Society of the Pacific Conference
  Series, Vol. 358, Solar Polarization 4, ed. R.~{Casini} \& B.~W. {Lites},
  107.
\newblock \doarXiv{astro-ph/0601483}

\bibitem[{{Borrero} \& {Ichimoto}(2011)}]{borrero11}
{Borrero}, J.~M., \& {Ichimoto}, K. 2011, Living Reviews in Solar Physics, 8,
  4, \dodoi{10.12942/lrsp-2011-4}

\bibitem[{{Borrero} \& {Kobel}(2011)}]{borrero11c}
{Borrero}, J.~M., \& {Kobel}, P. 2011, \aap, 527, A29,
  \dodoi{10.1051/0004-6361/201015634}

\bibitem[{{Borrero} {et~al.}(2019){Borrero}, {Pastor Yabar}, {Rempel}, \& {Ruiz
  Cobo}}]{borrero19}
{Borrero}, J.~M., {Pastor Yabar}, A., {Rempel}, M., \& {Ruiz Cobo}, B. 2019,
  \aap, 632, A111, \dodoi{10.1051/0004-6361/201936367}

\bibitem[{{Borrero} {et~al.}(2021){Borrero}, {Pastor Yabar}, \& {Ruiz
  Cobo}}]{borrero21}
{Borrero}, J.~M., {Pastor Yabar}, A., \& {Ruiz Cobo}, B. 2021, \aap, 647, A190,
  \dodoi{10.1051/0004-6361/202039927}

\bibitem[{Borrero {et~al.}(2011)Borrero, Tomczyk, Kubo, Socas-Navarro, Schou,
  Couvidat, \& Bogart}]{borrero11b}
Borrero, J.~M., Tomczyk, S., Kubo, M., {et~al.} 2011, Solar Physics, 273, 267

\bibitem[{Bueno(2010)}]{bueno10}
Bueno, J.~T. 2010, in Magnetic Coupling between the Interior and Atmosphere of
  the Sun, ed. S.~Hasan \& R.~J. Rutten (Berlin, Heidelberg: Springer Berlin
  Heidelberg), 118--140

\bibitem[{Casini {et~al.}(2009)Casini, Sainz, \& Low}]{casini09}
Casini, R., Sainz, R.~M., \& Low, B.~C. 2009, The Astrophysical Journal, 701,
  L43, \dodoi{10.1088/0004-637x/701/1/l43}

\bibitem[{{Centeno} {et~al.}(2014){Centeno}, {Schou}, {Hayashi}, {Norton},
  {Hoeksema}, {Liu}, {Leka}, \& {Barnes}}]{me_centeno14}
{Centeno}, R., {Schou}, J., {Hayashi}, K., {et~al.} 2014, \solphys, 289, 3531,
  \dodoi{10.1007/s11207-014-0497-7}

\bibitem[{Christensen-Dalsgaard {et~al.}(1990)Christensen-Dalsgaard, Schou, \&
  Thompson}]{dalsgaard1990}
Christensen-Dalsgaard, J., Schou, J., \& Thompson, M.~J. 1990, Monthly Notices
  of the Royal Astronomical Society, 242, 353, \dodoi{10.1093/mnras/242.3.353}

\bibitem[{{Christensen-Dalsgaard} \& {Thompson}(1993)}]{olarls1993}
{Christensen-Dalsgaard}, J., \& {Thompson}, M.~J. 1993, in Astronomical Society
  of the Pacific Conference Series, Vol.~42, GONG 1992. Seismic Investigation
  of the Sun and Stars, ed. T.~M. {Brown}, 249.
\newblock \url{https://ui.adsabs.harvard.edu/abs/1993ASPC...42..249C}

\bibitem[{{Collados} {et~al.}(1994){Collados}, {Martinez Pillet}, {Ruiz Cobo},
  {del Toro Iniesta}, \& {Vazquez}}]{collados94}
{Collados}, M., {Martinez Pillet}, V., {Ruiz Cobo}, B., {del Toro Iniesta},
  J.~C., \& {Vazquez}, M. 1994, \aap, 291, 622

\bibitem[{{de la Cruz Rodr{\'\i}guez} {et~al.}(2016){de la Cruz
  Rodr{\'\i}guez}, {Leenaarts}, \& {Asensio Ramos}}]{rodriguez16}
{de la Cruz Rodr{\'\i}guez}, J., {Leenaarts}, J., \& {Asensio Ramos}, A. 2016,
  \apjl, 830, L30, \dodoi{10.3847/2041-8205/830/2/L30}

\bibitem[{{de la Cruz Rodr{\'\i}guez} {et~al.}(2019){de la Cruz
  Rodr{\'\i}guez}, {Leenaarts}, {Danilovic}, \& {Uitenbroek}}]{stic19}
{de la Cruz Rodr{\'\i}guez}, J., {Leenaarts}, J., {Danilovic}, S., \&
  {Uitenbroek}, H. 2019, \aap, 623, A74, \dodoi{10.1051/0004-6361/201834464}

\bibitem[{{de la Cruz Rodr{\'\i}guez} \& {van Noort}(2017)}]{rodriguez17}
{de la Cruz Rodr{\'\i}guez}, J., \& {van Noort}, M. 2017, \ssr, 210, 109,
  \dodoi{10.1007/s11214-016-0294-8}

\bibitem[{del Toro~Iniesta(2003)}]{josecarlos_book}
del Toro~Iniesta, J.~C. 2003, Introduction to Spectropolarimetry (Cambridge
  University Press), \dodoi{10.1017/CBO9780511536250}

\bibitem[{{Del Toro Iniesta} \& {Ruiz Cobo}(1996)}]{sir96}
{Del Toro Iniesta}, J.~C., \& {Ruiz Cobo}, B. 1996, \solphys, 164, 169,
  \dodoi{10.1007/BF00146631}

\bibitem[{{del Toro Iniesta} \& {Ruiz Cobo}(2016)}]{sirLR16}
{del Toro Iniesta}, J.~C., \& {Ruiz Cobo}, B. 2016, Living Reviews in Solar
  Physics, 13, 4, \dodoi{10.1007/s41116-016-0005-2}

\bibitem[{{Frutiger} \& {Solanki}(1998)}]{frutiger98}
{Frutiger}, C., \& {Solanki}, S.~K. 1998, \aap, 336, L65

\bibitem[{{Frutiger} {et~al.}(2000){Frutiger}, {Solanki}, {Fligge}, \&
  {Bruls}}]{furtiger00}
{Frutiger}, C., {Solanki}, S.~K., {Fligge}, M., \& {Bruls}, J.~H.~M.~J. 2000,
  \aap, 358, 1109

\bibitem[{{Gingerich} {et~al.}(1971){Gingerich}, {Noyes}, {Kalkofen}, \&
  {Cuny}}]{hsra}
{Gingerich}, O., {Noyes}, R.~W., {Kalkofen}, W., \& {Cuny}, Y. 1971, \solphys,
  18, 347, \dodoi{10.1007/BF00149057}

\bibitem[{Golub \& Van~Loan(1996)}]{pinv2}
Golub, G.~H., \& Van~Loan, C.~F. 1996, {Matrix Computations}, 3rd edn. (The
  Johns Hopkins University Press, Baltimore, MD).
\newblock
  \url{https://twiki.cern.ch/twiki/pub/Main/AVFedotovHowToRootTDecompQRH/Golub_VanLoan.Matr_comp_3ed.pdf}

\bibitem[{Gough(1985)}]{gough1985}
Gough, D. 1985, Solar Physics, 100, 65, \dodoi{10.1007/BF00158422}

\bibitem[{Gough(1982)}]{gough1982}
Gough, D.~O. 1982, Nature, 298, 334, \dodoi{10.1038/298334a0}

\bibitem[{Gray(2005)}]{gray2005}
Gray, D.~F. 2005, The Observation and Analysis of Stellar Photospheres, 3rd
  edn. (Cambridge University Press), \dodoi{10.1017/CBO9781316036570}

\bibitem[{Hansen(1990)}]{hansen_dpc2}
Hansen, P.~C. 1990, SIAM Journal on Scientific and Statistical Computing, 11,
  503, \dodoi{10.1137/0911028}

\bibitem[{Hansen(1994)}]{hansen94}
---. 1994, Numerical Algorithms, 6, 1

\bibitem[{Hansen(1998)}]{hansen_rls2}
---. 1998, Rank-deficient and discrete ill-posed problems: numerical aspects of
  linear inversion (SIAM)

\bibitem[{{Harvey} {et~al.}(1972){Harvey}, {Livingston}, \&
  {Slaughter}}]{harvey72}
{Harvey}, J., {Livingston}, W., \& {Slaughter}, C. 1972, in Line Formation in
  the Presence of Magnetic Fields, 227

\bibitem[{{Jeffrey}(1988)}]{jeffrey1988}
{Jeffrey}, W. 1988, \apj, 327, 987, \dodoi{10.1086/166255}

\bibitem[{{Landi Degl'Innocenti} \& {Landi Degl'Innocenti}(1977)}]{landi77}
{Landi Degl'Innocenti}, E., \& {Landi Degl'Innocenti}, M. 1977, \aap, 56, 111

\bibitem[{Landolfi {et~al.}(1984)Landolfi, Landi~Degl'innocenti, \&
  Arena}]{landolfi84}
Landolfi, M., Landi~Degl'innocenti, E., \& Arena, P. 1984, Solar Physics, 93,
  269

\bibitem[{{Lites} {et~al.}(2008){Lites}, {Kubo}, {Socas-Navarro}, {Berger},
  {Frank}, {Shine}, {Tarbell}, {Title}, {Ichimoto}, {Katsukawa}, {Tsuneta},
  {Suematsu}, {Shimizu}, \& {Nagata}}]{lites08}
{Lites}, B.~W., {Kubo}, M., {Socas-Navarro}, H., {et~al.} 2008, \apj, 672,
  1237, \dodoi{10.1086/522922}

\bibitem[{{Mart{\'\i}nez Gonz{\'a}lez} {et~al.}(2008){Mart{\'\i}nez
  Gonz{\'a}lez}, {Asensio Ramos}, {Carroll}, {Kopf}, {Ram{\'\i}rez V{\'e}lez},
  \& {Semel}}]{gonzalez08}
{Mart{\'\i}nez Gonz{\'a}lez}, M.~J., {Asensio Ramos}, A., {Carroll}, T.~A.,
  {et~al.} 2008, \aap, 486, 637, \dodoi{10.1051/0004-6361:200809719}

\bibitem[{{Mihalas}(1978)}]{mihalas78}
{Mihalas}, D. 1978, {Stellar atmospheres}.
\newblock \url{https://ui.adsabs.harvard.edu/abs/1978stat.book.....M}

\bibitem[{{Mili{\'c}} \& {van Noort}(2017)}]{milicRF}
{Mili{\'c}}, I., \& {van Noort}, M. 2017, \aap, 601, A100,
  \dodoi{10.1051/0004-6361/201629980}

\bibitem[{{Mili{\'c}} \& {van Noort}(2018)}]{milic18}
---. 2018, \aap, 617, A24, \dodoi{10.1051/0004-6361/201833382}

\bibitem[{{Morosin} {et~al.}(2022){Morosin}, {de la Cruz Rodr{\'\i}guez},
  {D{\'\i}az Baso}, \& {Leenaarts}}]{morosin22}
{Morosin}, R., {de la Cruz Rodr{\'\i}guez}, J., {D{\'\i}az Baso}, C.~J., \&
  {Leenaarts}, J. 2022, \aap, 664, A8, \dodoi{10.1051/0004-6361/202243461}

\bibitem[{{Orozco Su{\'a}rez} \& {Del Toro Iniesta}(2007)}]{suarez07a}
{Orozco Su{\'a}rez}, D., \& {Del Toro Iniesta}, J.~C. 2007, \aap, 462, 1137,
  \dodoi{10.1051/0004-6361:20066201}

\bibitem[{{Orozco Suarez} {et~al.}(2005){Orozco Suarez}, {Lagg}, \&
  {Solanki}}]{suarez05}
{Orozco Suarez}, D., {Lagg}, A., \& {Solanki}, S.~K. 2005, in ESA Special
  Publication, Vol. 596, Chromospheric and Coronal Magnetic Fields, ed. D.~E.
  {Innes}, A.~{Lagg}, \& S.~A. {Solanki}, 59.1

\bibitem[{{Orozco Su{\'a}rez} {et~al.}(2007){Orozco Su{\'a}rez}, {Bellot
  Rubio}, {del Toro Iniesta}, {Tsuneta}, {Lites}, {Ichimoto}, {Katsukawa},
  {Nagata}, {Shimizu}, {Shine}, {Suematsu}, {Tarbell}, \& {Title}}]{suarez07}
{Orozco Su{\'a}rez}, D., {Bellot Rubio}, L.~R., {del Toro Iniesta}, J.~C.,
  {et~al.} 2007, \apjl, 670, L61, \dodoi{10.1086/524139}

\bibitem[{{Pijpers} \& {Thompson}(1992)}]{pijper92}
{Pijpers}, F.~P., \& {Thompson}, M.~J. 1992, \aap, 262, L33.
\newblock \url{https://ui.adsabs.harvard.edu/abs/1992A&A...262L..33P}

\bibitem[{{Pijpers} \& {Thompson}(1994)}]{thompson94}
---. 1994, \aap, 281, 231.
\newblock \url{https://ui.adsabs.harvard.edu/abs/1994A&A...281..231P}

\bibitem[{Press {et~al.}(2007)Press, Teukolsky, Vetterling, \&
  Flannery}]{numrecipe}
Press, W.~H., Teukolsky, S.~A., Vetterling, W.~T., \& Flannery, B.~P. 2007,
  Numerical Recipes 3rd Edition: The Art of Scientific Computing, 3rd edn.
  (USA: Cambridge University Press).
\newblock
  \url{https://www.cambridge.org/us/academic/subjects/mathematics/numerical-recipes/numerical-recipes-art-scientific-computing-3rd-edition?format=HB&isbn=9780521880688}

\bibitem[{{Priest} {et~al.}(2018){Priest}, {Chitta}, \& {Syntelis}}]{priest18}
{Priest}, E.~R., {Chitta}, L.~P., \& {Syntelis}, P. 2018, \apjl, 862, L24,
  \dodoi{10.3847/2041-8213/aad4fc}

\bibitem[{Ramos {et~al.}(2012)Ramos, Sainz, Gonz{\'{a}}lez, Viticchi{\'{e}},
  Su{\'{a}}rez, \& Socas-Navarro}]{ramos12}
Ramos, A.~A., Sainz, R.~M., Gonz{\'{a}}lez, M. J.~M., {et~al.} 2012, The
  Astrophysical Journal, 748, 83, \dodoi{10.1088/0004-637x/748/2/83}

\bibitem[{Rempel(2014)}]{rempel14}
Rempel, M. 2014, The Astrophysical Journal, 789, 132,
  \dodoi{10.1088/0004-637x/789/2/132}

\bibitem[{{Ruiz Cobo} \& {del Toro Iniesta}(1992)}]{sir92}
{Ruiz Cobo}, B., \& {del Toro Iniesta}, J.~C. 1992, \apj, 398, 375,
  \dodoi{10.1086/171862}

\bibitem[{{Ruiz Cobo} {et~al.}(2022){Ruiz Cobo}, {Quintero Noda}, {Gafeira},
  {Uitenbroek}, {Orozco Su{\'a}rez}, \& {P{\'a}ez Ma{\~n}{\'a}}}]{desire22}
{Ruiz Cobo}, B., {Quintero Noda}, C., {Gafeira}, R., {et~al.} 2022, \aap, 660,
  A37, \dodoi{10.1051/0004-6361/202140877}

\bibitem[{{Skumanich} \& {Lites}(1985)}]{skumanich85}
{Skumanich}, A., \& {Lites}, B.~W. 1985, in Measurements of Solar Vector
  Magnetic Fields, ed. M.~J. {Hagyard}, 341

\bibitem[{{Skumanich} \& {Lites}(1987)}]{me_sku87}
{Skumanich}, A., \& {Lites}, B.~W. 1987, \apj, 322, 473, \dodoi{10.1086/165743}

\bibitem[{{Socas-Navarro}(2001)}]{navarro01}
{Socas-Navarro}, H. 2001, in Astronomical Society of the Pacific Conference
  Series, Vol. 236, Advanced Solar Polarimetry -- Theory, Observation, and
  Instrumentation, ed. M.~{Sigwarth}, 487

\bibitem[{{Socas-Navarro} {et~al.}(2015){Socas-Navarro}, {de la Cruz
  Rodr{\'\i}guez}, {Asensio Ramos}, {Trujillo Bueno}, \& {Ruiz
  Cobo}}]{navarro15}
{Socas-Navarro}, H., {de la Cruz Rodr{\'\i}guez}, J., {Asensio Ramos}, A.,
  {Trujillo Bueno}, J., \& {Ruiz Cobo}, B. 2015, \aap, 577, A7,
  \dodoi{10.1051/0004-6361/201424860}

\bibitem[{{Socas-Navarro} {et~al.}(2000){Socas-Navarro}, {Trujillo Bueno}, \&
  {Ruiz Cobo}}]{navarro00}
{Socas-Navarro}, H., {Trujillo Bueno}, J., \& {Ruiz Cobo}, B. 2000, \apj, 530,
  977, \dodoi{10.1086/308414}

\bibitem[{{Solanki}(2003)}]{solanki03}
{Solanki}, S.~K. 2003, \aapr, 11, 153, \dodoi{10.1007/s00159-003-0018-4}

\bibitem[{{Solanki} {et~al.}(1992){Solanki}, {Rueedi}, \&
  {Livingston}}]{solanki92}
{Solanki}, S.~K., {Rueedi}, I.~K., \& {Livingston}, W. 1992, \aap, 263, 312

\bibitem[{Stenflo(1982)}]{stenflo82}
Stenflo, J.~O. 1982, Solar Physics, 80, 209

\bibitem[{{Stenflo}(2010)}]{stenflo10}
{Stenflo}, J.~O. 2010, \aap, 517, A37, \dodoi{10.1051/0004-6361/200913972}

\bibitem[{{Trelles Arjona} {et~al.}(2021){Trelles Arjona}, {Mart{\'\i}nez
  Gonz{\'a}lez}, \& {Ruiz Cobo}}]{arjona21}
{Trelles Arjona}, J.~C., {Mart{\'\i}nez Gonz{\'a}lez}, M.~J., \& {Ruiz Cobo},
  B. 2021, \apjl, 915, L20, \dodoi{10.3847/2041-8213/ac0af2}

\bibitem[{{Trujillo Bueno} {et~al.}(2004){Trujillo Bueno}, {Shchukina}, \&
  {Asensio Ramos}}]{bueno04}
{Trujillo Bueno}, J., {Shchukina}, N., \& {Asensio Ramos}, A. 2004, \nat, 430,
  326, \dodoi{10.1038/nature02669}

\bibitem[{{Unno}(1956)}]{me_unno56}
{Unno}, W. 1956, \pasj, 8, 108

\bibitem[{{V{\"o}gler} {et~al.}(2005){V{\"o}gler}, {Shelyag}, {Sch{\"u}ssler},
  {Cattaneo}, {Emonet}, \& {Linde}}]{voegler05}
{V{\"o}gler}, A., {Shelyag}, S., {Sch{\"u}ssler}, M., {et~al.} 2005, \aap, 429,
  335, \dodoi{10.1051/0004-6361:20041507}

\bibitem[{{Yadav} {et~al.}(2020){Yadav}, {Cameron}, \& {Solanki}}]{yadav20}
{Yadav}, N., {Cameron}, R.~H., \& {Solanki}, S.~K. 2020, \apjl, 894, L17,
  \dodoi{10.3847/2041-8213/ab8dc5}

\bibitem[{Yadav {et~al.}(2017)Yadav, Mathew, \& Tiwary}]{Yadav2017}
Yadav, R., Mathew, S.~K., \& Tiwary, A.~R. 2017, Solar Physics, 292, 105,
  \dodoi{10.1007/s11207-017-1131-2}

\end{thebibliography}
\bibliographystyle{aasjournal}

\appendix

\section{Spectral lines list}

\begin{table}[!h]
	\centering
	 \begin{tabular}{c c} 
	 \hline
	$\lambda_0$ [\AA] & Blends (if any)   \\ [0.5ex]  
	 \hline\hline \\
	 Fe II (6147.743) & Fe I (6147.835)  \\ [0.5ex] 
	 Fe II (6149.241) & -                \\ [0.5ex] 
	 Ti I  (6149.725) & -                \\ [0.5ex] 
	 Fe II (6150.113) & V I (6150.167)   \\ [0.5ex] 
	 Fe I  (6301.499) & Fe I (6302.493)  \\ [0.5ex] 
	 Fe I  (15648.515) & Fe I (15647.423) \\ [0.5ex] 
	 Fe I  (15652.882) & -                \\ [0.5ex] 
	 \hline
	 \end{tabular}
	 \caption{List of spectral lines used in this work.} 
	  \label{table_linelist}
\end{table}


\section{SIR averaging kernels}
\label{Appendix_sirkernel}

When solving, say Equation~\ref{eqn2_1storder1var}, SIR inverts an overdetermined version of \fRFTm, to prevent \errv\ amplification. The goal is first to compute \fdTvflat\ solutions at a limited $l$ node locations, and finally interpolate nodal values to compute solutions at all $m$ depth points. 
Overdetermined \fRFTm, computed with the help of interpolation coefficient matrix \Fm, is given by \Fm $\bigcdot$ \fRFTm, and contains 'equivalent' response function sensitvity to node locations \cite[page 209]{josecarlos_book}. 

Nodal \fdTvflat\ solutions, using the inverse of equivalent response functions, are given by 
		\begin{eqnarray} 
			[( \mathbf{F} \bigcdot\ \mathbf{R}_{\text{T}})^\top]^{-1} \bigcdot \Delta \mathbf{I}\ &=& \  [( \mathbf{F} \bigcdot\ \mathbf{R}_{\text{T}})^\top]^{-1} \bigcdot [\mathbf{R}_\text{T}^\top \bigcdot \frac{\Delta{\mathbf{T}}}{\mathbf{T}} \  + \  \boldsymbol{\varepsilon}] . 
		\end{eqnarray}
\noindent
From these $l$ nodal values, SIR inversion solution at all $m$ depths is given by

		\begin{eqnarray} 
			\mathbf{F}^\top \bigcdot [( \mathbf{F} \bigcdot\ \mathbf{R}_{\text{T}})^\top]^{-1} \bigcdot \Delta \mathbf{I}\ &=& \  \mathbf{F}^\top \bigcdot [( \mathbf{F} \bigcdot\ \mathbf{R}_{\text{T}})^\top]^{-1} \bigcdot [\mathbf{R}_\text{T}^\top \bigcdot \frac{\Delta{\mathbf{T}}}{\mathbf{T}} \  + \  \boldsymbol{\varepsilon}] .
		\end{eqnarray}

\noindent
Here, individual rows of matrix \Fm $^\top \bigcdot[($\Fm $\bigcdot$ \fRFTm$)^{\top}]^{-1} \bigcdot$ \fRFTmt\ correspond to SIR averaging kernels at each depth locations. \\ 


\section{SIR inversion parameters}
\label{appendix_sir}

\begin{table}[!h]
	\centering
		 \begin{tabular}{c c c c c} 
			 \hline
			Inverting & T nodes & \Vlos\ nodes  & \Pe\ nodes  & Number of cycles  \\ [0.5ex] 
			 \hline\hline
			 T & 1,2,3,4,5,6,8,11$^a$ & 0 & 1$^b$ & 8   \\ [1ex] 
			 T, \Vlos\ and \Pe\ & 1,2,3,4,5,6,8,11 & 1,2,3,4,5,6,8,11 & 1,2,3,4,5$^b$ & 8  \\ [1ex]
			 \hline
		 \end{tabular} 
		 \caption{SIR node values}
	 \footnotesize 
	 	\begin{flushleft}
	 		\textbf{Notes.} \\
	 		$^a$ Each comma separated values correspond to the node used in a given cycle. \\
	 		$^b$ If nodes for \Pe\ inversion is set to 0, then SIR inverts for T assuming {hydrostatic equilibrium}. To be consistent with single variable T inversion, SIR inverted \dPev\ is manually set to 0. \\
	 		$^c$ SIR inverted \Pe\ solutions (generally) deviate drastically from underlying \Pe\ model for larger node values. 
 		\end{flushleft}	

	  \label{table2}
\end{table}

\end{document}